\documentclass{aastex631}

\usepackage{amsmath}
\usepackage{multirow}
\usepackage{color}

\shorttitle{Long-term optical variability of S5~0716+714}
\shortauthors{Gorbachev et al.}

\graphicspath{{./}{figures/}}

\received{September 10, 2021}
\revised{December 27, 2021}
\accepted{January 19, 2022}
\submitjournal{ApJ}

\begin{document}

\title{Long-term multiband optical variability of S5~0716+714 blazar}

\author[0000-0001-5731-9264]{Mark A. Gorbachev}
\affiliation{Crimean Astrophysical Observatory, Nauchny 298409, Crimea, Russia}

\author[0000-0001-7307-2193]{Marina S. Butuzova}
\affiliation{Crimean Astrophysical Observatory, Nauchny 298409, Crimea, Russia}

\author{Sergey G. Sergeev}
\affiliation{Crimean Astrophysical Observatory, Nauchny 298409, Crimea, Russia}

\author{Sergey V. Nazarov}
\affiliation{Crimean Astrophysical Observatory, Nauchny 298409, Crimea, Russia}

\author{Alexey V. Zhovtan}
\affiliation{Crimean Astrophysical Observatory, Nauchny 298409, Crimea, Russia}

\begin{abstract}

Multiband optical photometry data of blazar S5~0716+714 obtained from 2002 to 2019 reveal stable color index change with flux variability. 
We analyzed this trend under variability caused by the Doppler factor change in the presence of a curved photon energy spectrum.  
A break in the energy spectrum of emitting electrons, caused by radiative losses, or log-parabolic electron energy distribution, or the synchrotron self-absorption acting in a compact jet part forms such the photon spectrum.
We explained the observed color index change with variability by geometric effects only under the assumption that the radiating region is the synchrotron self-absorbed core and the bright optically thin jet. 
In this framework, we estimated the magnetic field strength in the optically thick part of the radiating region. These values correspond to other independent estimates of the magnetic field near the black hole, further supporting our assumption.

\end{abstract}


\section{Introduction} \label{sec:Introduction}

S5~0716+714 is a bright blazar, convenient for observation in the northern hemisphere, having violent variability across all observed electromagnetic spectrum. 
In different periods of observation, the object exhibits different color properties of the long-term variability. 
For example, the \citet{Ghisellini97,Raiteri03,Volvach12} reported that the color index  does not change under optical flux variability.
According to other data, moderate or strong chromatism manifested: the object color becomes bluer with  brightness increase \citep{Wu07, Stalin09, Zhang18, Kaur18}.
As \citet{Bhatta13, Hong17} noted, the chromatism does not depend on the object magnitude but most likely depends on the activity state: the object is bluer when it is more active. 
\citet{Wu05,Ghisellini97} detected bluer-when-brighter (BWB) trend during optical flares, while BWB was absent in relatively quiet states. 

The analysis of color change with flux variability is important because it allows us to conclude the variability mechanism.
There is a common assertion. If the variability is achromatic, it is caused by geometric effects \citep{CamKrock92}, otherwise, the change in the object brightness has physical causes \citep[e.g., shocks in a jet,][]{MarscherGear85}.
It is true if the energy spectrum of the emitting electrons is power-law and the radiation comes from an optically thin medium.
If the electron energy spectrum flattens at low energies or the medium is optically thick for radiation at lower frequencies, then an increase in the Doppler factor ($\delta$) leads to radiation characterized by a flatter spectrum in the observed range.
Then the BWB chromatism is observed.

An alternative interpretation of BWB chromatism is that the two components having ``red'' and ``blue'' colors contribute to the observed radiation flux.
At first, it was assumed that the red component represents the constant radiation of the underlying galaxy, while the blue one is the radiation formed in a relativistic jet \citep{Sandage73}.
A more recent interpretation is that the red component is associated with the synchrotron radiation of the jet, while the blue one is associated with the thermal radiation of the accretion disc \citep[see, e.g.,][and references therein]{GuLee06, Isler17}.
Then the observed BWB chromatism originates due to a change in the contribution of the red and blue components to the total flux.
These interpretations of the object's color index variability aren't suitable for BL~Lac objects since their spectrum is quasi-featureless, excluding the accretion disk contribution to the observed flux.

In this paper, using the observational data from 2002 to 2019, we analyze the long-term multiband optical variability of the blazar S5~0716+714  (Section~\ref{sec:Optical}).
In Section~3, we determine the variable component spectrum based on observational data and show that the radiation of the host galaxy as a constant component in the radiation is not applicable for this object.
% This result is 
This result is expected because the observed non-thermal continuum spectrum of S5~0716+714 without any spectral line \citep{Nilsson08} indicates the radiation formation in the ultrarelativistic jet. 
As the cavity in the broadband spectral energy distribution of the blazar occurs in the X-ray range \citep{Liao14}, then the optical radiation is synchrotron.
Within this framework, we make various assumptions on the radiating region and evaluate the possibility of interpreting the observed characteristics of long-term variability only by geometric effects for the non-power-law emission spectrum.
In Section~4, we consider synchrotron self-absorption as the cause of the curved photon spectrum. 
Under this assumption, we estimate the magnetic field strength and minimum Lorentz factor of electron energy distribution in the optical emission region (Section~5). The broken power-law and log-parabolic electron energy spectra are examined in Section~6.
Sections~7 and 8 contain a discussion of the obtained results and conclusions, respectively.

\section{Optical observations and data analysis} \label{sec:Optical}

\subsection{Observations and light curve} \label{subsec:Observations}

\begin{figure}
\plotone{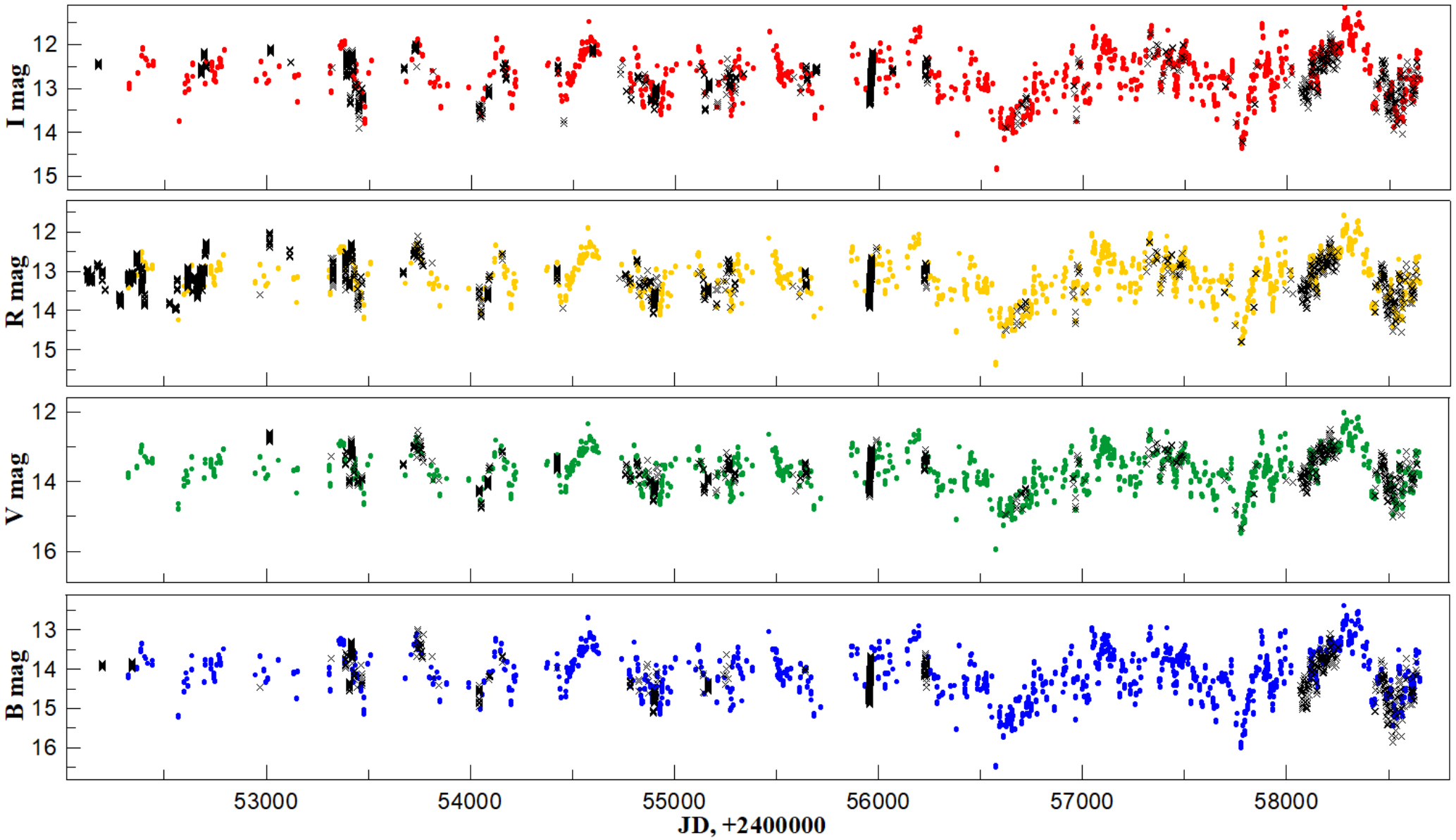}
\caption{The S5~0716+714 light curves  in the B, V, R, I bands (from top to bottom, respectively). Colors and crosses mark the AZT-8 data and data from other sources (see references in the text), respectively.} 
\label{fig:lightcurve}
\end{figure}

The observational data have been obtained on the 70-cm AZT-8 telescope of the Crimean Astrophysical Observatory of the Russian Academy of Sciences from 02.2002 to 06.2019.
The observations were carried out using broadband filters B, V, R, and I of the Johnson system.
The instrumentation, processing, and measurements of our photometric data are described by \citet{Doroshenko2005} and \citet{Sergeev2005}.
The interstellar absorption in the S5~0716+714 direction was accounted for with the coefficients from \citet{Ext}.
As photometric standards, we used the stars designated 8, 11, and 18 in~\citet{Doroshenko2005}, which \citet{Villata1998} mark as 3, 5, and 6, respectively.
Figure~\ref{fig:lightcurve} shows the light curve; the measurement errors do not exceed the symbol size. 
The black crosses mark other published data  \citep{Raiteri03, Dai15, Ghisellini97, Hong17, Poon09, Liao14, Zhang08, Bhatta13, Wu12, Wu07, Feng20, Xu2019, Xiong2020, Liu2019}.
Our observations cover different periods of the object's activity.
In JD~2456573.3, the maximum magnitudes $m_\text{B}=16.49$, $m_\text{V}=15.96$, $m_\text{R}=15.37$, and $m_\text{I}=14.84$ are registered in the B, V, R, and I bands, respectively.
The minimum magnitudes of $m_\text{B}=12.37$, $m_\text{V}=12.01$, $m_\text{R}=11.56$, and $m_\text{I}=11.15$
was observed in JD~2458279.5.
The AZT-8 data represents the most continuous multiband data set with approximately a similar time resolution.

\subsection{Two component variability model} \label{subsec:2.2}

\begin{figure}
\centering
\includegraphics[scale=0.6]{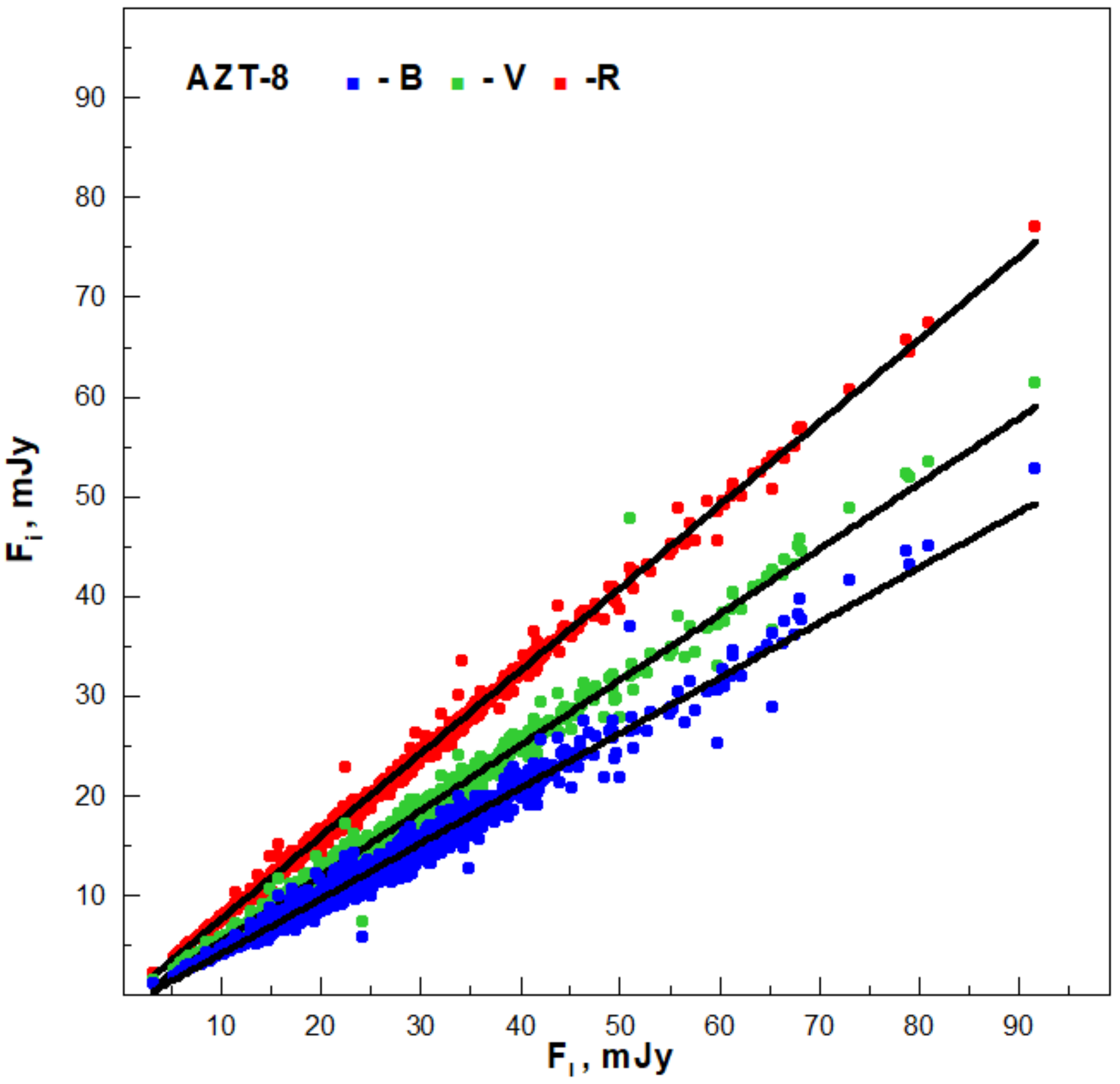} 
\caption{The flux-flux diagrams based on AZT-8 observation data. Colored markers indicate the B, V, R fluxes relative to $F_\text{I}$. The lines show linear approximations of data.} 
\label{fig:FF_all}
\end{figure}

\begin{figure}
\plotone{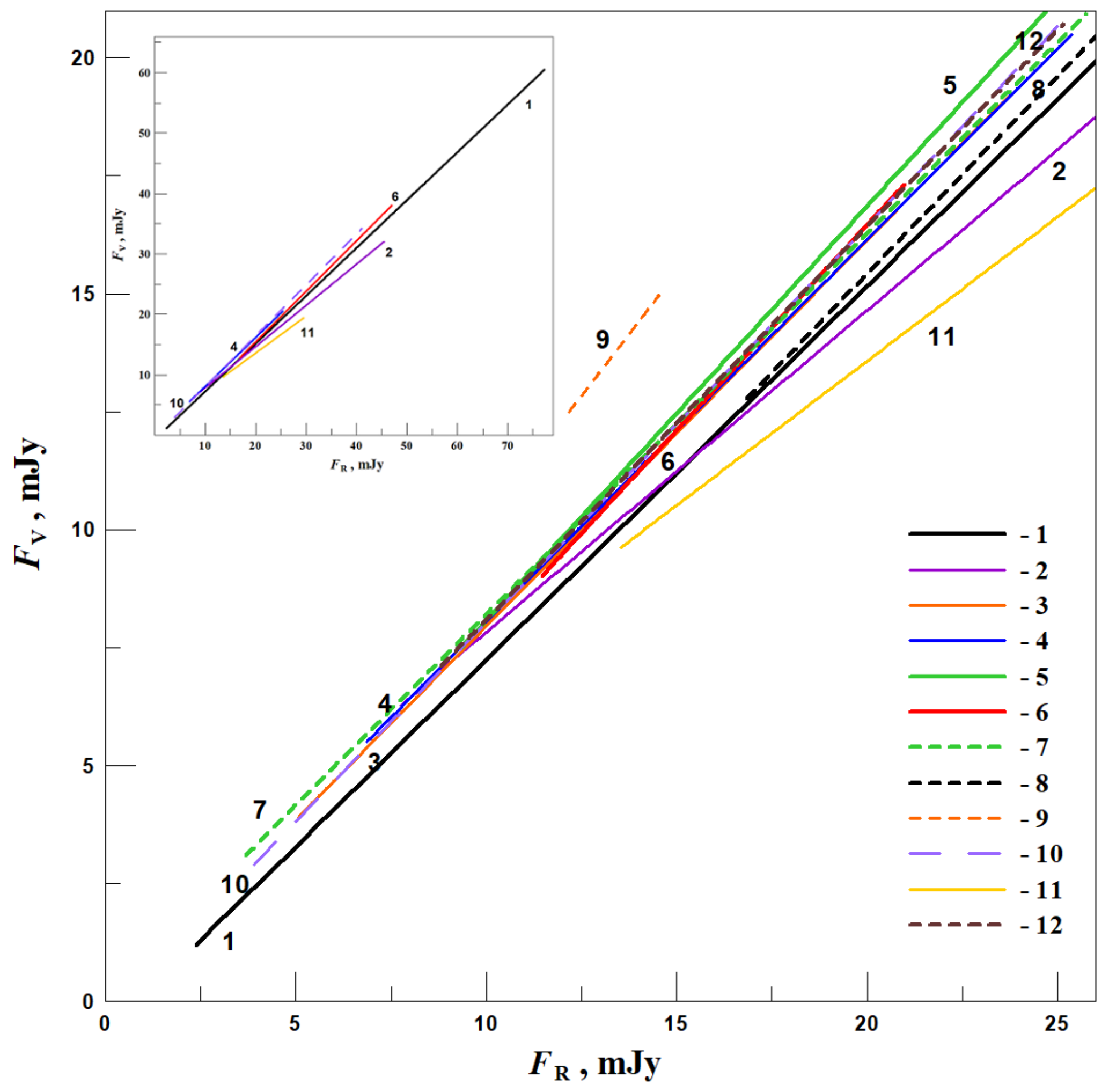}
\caption{The flux-flux diagrams in the R and B bands for region $F_{\text{R}}<25$~mJy. Colored lines show linear fits of data from various sources: 1 --- AZT-8 from 02.2002 to 06.2019, 2 --- \citet{Dai15} from 01.2005 to 11.2012, 3 --- \citet{Feng20} from 11.2018 to 03.2019, 4 --- \citet{Ghisellini97} from 11.1994 to 04.1996, 5 --- \citet{Hong17} from 01.2012 to 02.2012, 6 --- \citet{Liao14} from 09.2008 to 04.2011, 7 --- \citet{Poon09} from 10.2008 to 02.2009, 8 --- \citet{Raiteri03} from 11.1994 to 01.20027, 9 --- \citet{Wu07,Wu12} from 01.2006 to 02.2006 and 12.2006, 10 --- \citet{Xiong2020} from 11.2017 to 06.2019, 11 --- \citet{Xu2019} from 01.2011 to 02.2018, 12 --- \citet{Zhang08} from 02.2001 to 04.2006. We plot the full range of $F_\text{R}$ changes in the upper-left corner.} 
\label{fig:FF_RV_all}
\end{figure}

The changes in brightness are often analyzed within the framework of a two-component model, containing constant and variable components.
These components have different colors, and their relative contribution to the total observed flux determines the observed object color index.
As S5~0716+714 shows the BWB trend, therefore the variable component has a blue color.
However, it is not clear whether the variable component color changes depending on its brightness.
To answer this question, \citet{Choloniewski81}, later \citet{Hagen-Thorn97, Larionov2008} used the flux-flux diagrams.
To plot the diagrams for S5~0716+714 (Fig.~\ref{fig:FF_all}), we used the absolute calibration formulae by \citet{Mead90} for conversion magnitudes to flux densities.
The lines show the linear fit (further in the text, we call it ``the observed lines'') of the observational data for three filter pairs.
Analytical expressions of the observed lines is
\begin{eqnarray}
\label{eq:obs_line}
F_{\text{B}} &= 0.553 \cdot F_{\text{I}} - 1.356, \nonumber \\
F_{\text{V}} &= 0.656 \cdot F_{\text{I}} - 1.151, \\
F_{\text{R}} &= 0.829 \cdot F_{\text{I}} - 0.636. \nonumber
\end{eqnarray}
The approximations by the linear function have a high Pearson correlation coefficient ($\geq0.975$), so we consider it unreasonable to fit the observed dependences by a polynomial.
Figure~\ref{fig:FF_RV_all} displays the observed lines on the flux-flux diagram for a pair of R---V filters according to AZT-8 data,~\citet{Dai15, Ghisellini97, Hong17, Liao14, Poon09, Wu07, Wu12, Zhang08, Raiteri03, Feng20, Xu2019, Xiong2020}.
The Figure~\ref{fig:FF_RV_all} caption contains the time intervals considered in these references.
We have reduced the displayed range of $F_\text{R}$ and $F_\text{B}$ for better visibility since only AZT-8 data has large flux values.
Figure~\ref{fig:FF_RV_all} shows that the various observed lines are consistent and do not escape beyond the AZT-8 point spread.
The observed lines plotted from \citet{Raiteri03} and AZT-8 data agree well, although \citet{Raiteri03} observed for a long time before the beginning of the AZT-8 ones.
We combined data of \citet{Wu07} and~\citet{Wu12} into one set because the observations were made on the same telescope but at different time intervals (01.01.2006---01.02.2006 and 18.12.2006---27.12.2006, respectively).
S5~0716+714 has steady long-term dependencies of fluxes in the different pairs of optical bands.

\citet{Hagen-Thorn97} supposed that the point corresponding to the constant component fluxes lies in the extrapolation of the observed line to small flux values.
This assumption is attractive because, in this case, a BWB trend exists under the constant spectrum of the variable component.
\citet{Hagen-Thorn97} noted that if the constant component point deviates from the extrapolation line, then the variable component spectrum changes in a complex manner.
They consider this change hard to explain physically.

\begin{figure}
\plotone{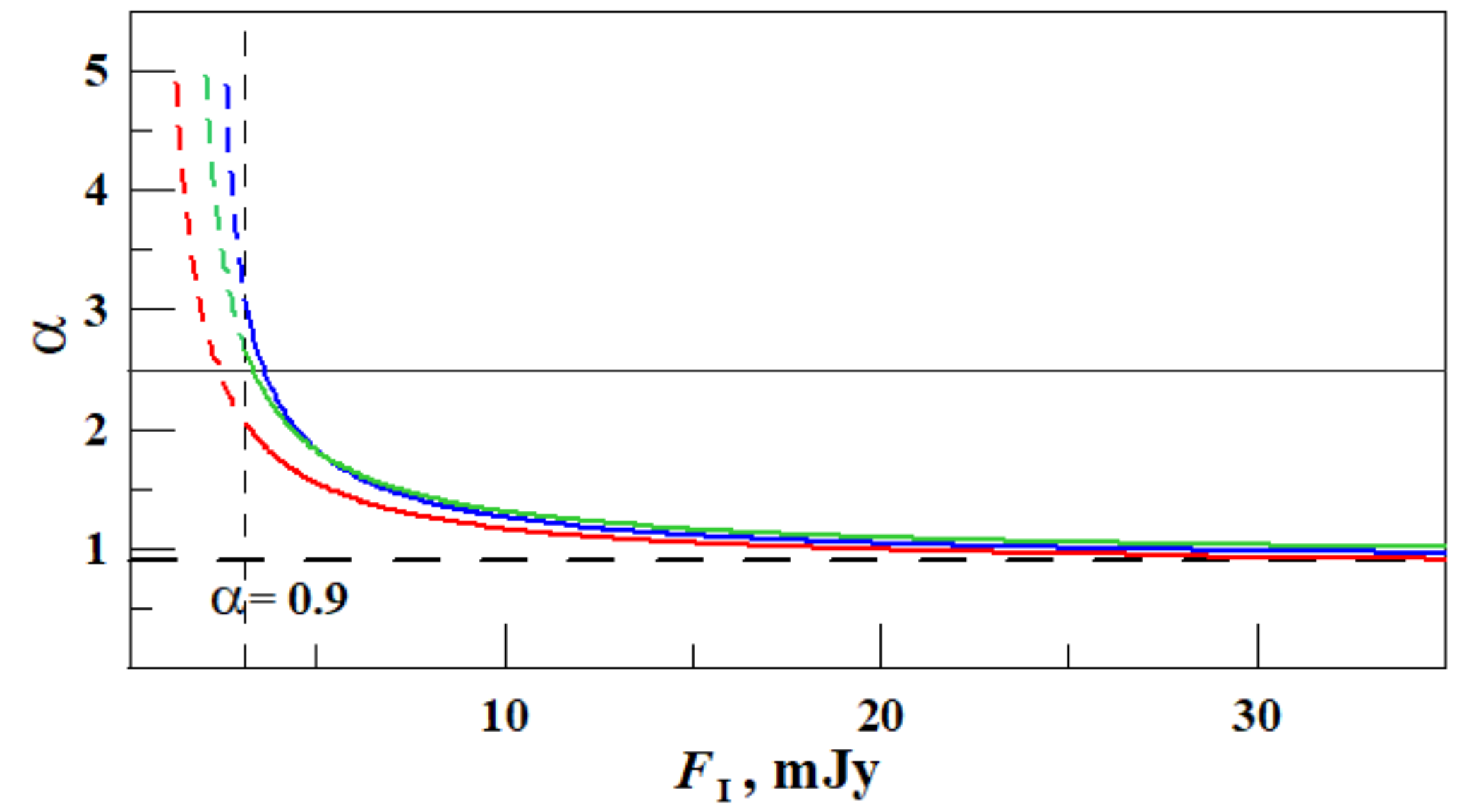}
\caption{The spectral index depending on $F_\text{I}$, calculated from the observed lines and their extrapolation to low values of $F_\text{I}$. The blue, green, and red colors mark the B --- I, V --- I, and R --- I band pairs, respectively. Solid lines refer to the range of observed values of $F_\text{I}$, dotted lines is to non-detectable small $F_\text{I}$ values. The horizontal dashed line corresponds to the asymptotic value of the spectral index defined for all band pairs. A solid horizontal line marks the spectral index maximum possible value of 2.5 for synchrotron radiation of electrons, which have experienced significant energy losses. The vertical line shows the minimum registered flux $F_\text{I}$.} 
\label{fig:alpha_obs}
\end{figure}

We analyzed the available data to identify the constant component in radiation.
Based on the expressions (\ref{eq:obs_line}), we constructed the dependence of the spectral index on $F_{\text{I}}$ (Fig.~\ref{fig:alpha_obs}).
There is the BWB trend, which is strongly manifested at small values of $F_{\text{I}}$ and is insignificant at bright states of the object.
\citet{Feng20} obtained a similar result.
If we assume that the constant component point locates on the extrapolation of the observed lines at some fixed $F_{\text{I}}$, then its spectrum is not power-law and very steep.
It corresponds neither to the host elliptical galaxy or the synchrotron source spectrum.
This result is quite understandable since, e. g., \citet{Pursimo02, Wagner96, Sbarufatti05} didn't detect the underlying galaxy. Therefore, all observed radiation forms in the jet.
Then we can assume that in the low state, the constant component radiation significantly dominates in the object's total flux. 
Small flux values are rare (data with $F_\text{I}\leq 10$~mJy was about $6\%$ of all points and appeared for a short time interval).
For the flux of $F_\text{I}\leq 10$~mJy, the object spectrum is steep (Fig.~\ref{fig:alpha_obs}), and it can be associated with a constant component, whose emitting electrons were strongly cooled. 
Additionally, the contribution of the variable component to the total flux of the object in the low state is negligible.
This object state has been observed rarely and seems atypical, so we believe that $F_\text{I}^\text{c}=10$~mJy represents an acceptable estimate of the constant component flux.

\begin{figure}
\plotone{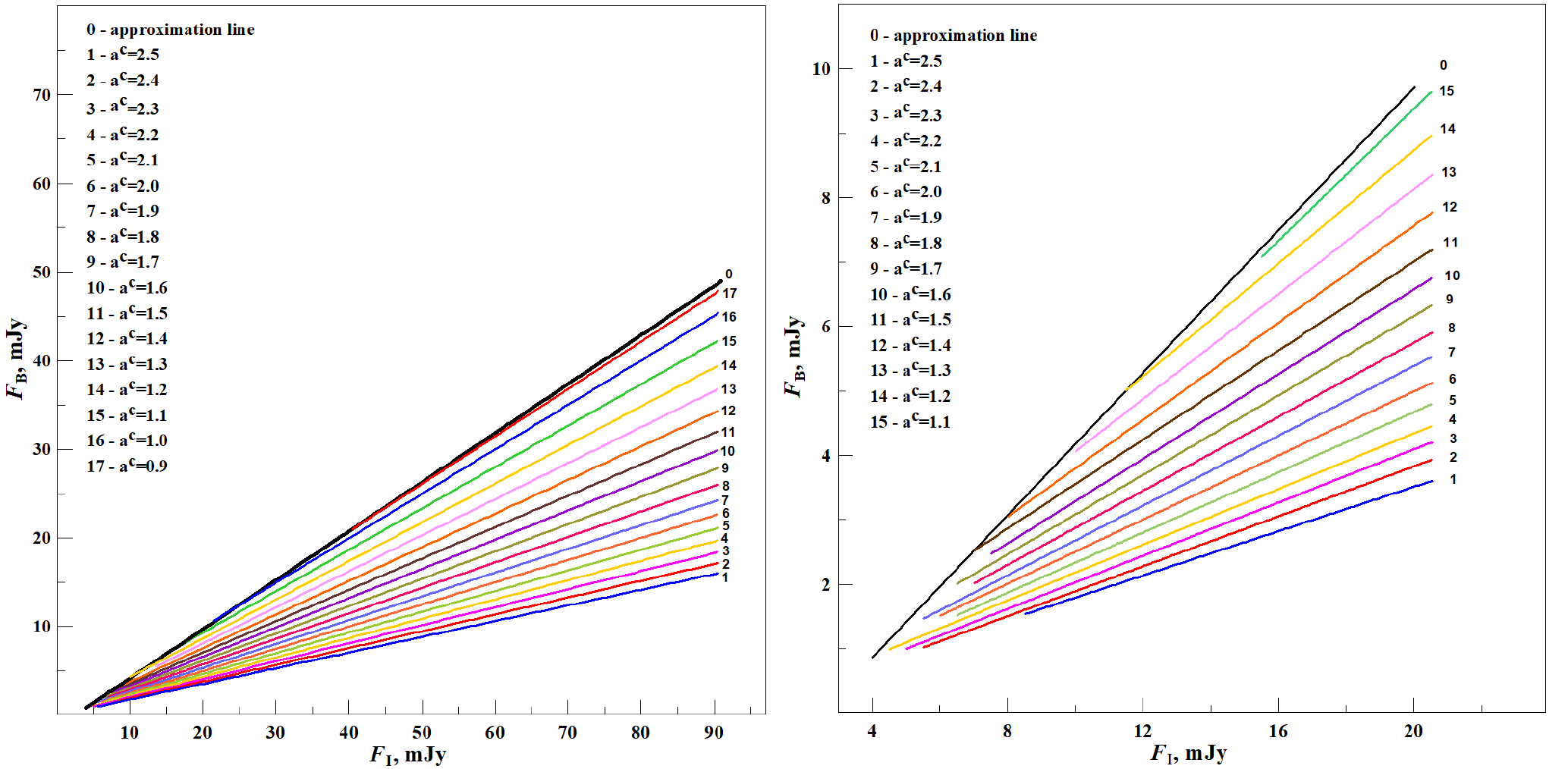}
\caption{Distribution of spectral indices on the flux-flux diagram. Colored lines represent points with the corresponding $\alpha_\text{c}=\text{const}$. The black line marks the observed line. The left panel shows the full range of changes in $F_\text{I}$, the right panel shows a section with $F_\text{I}<20$~mJy.} 
\label{fig:A_const}
\end{figure}

If we assume that the maximum value of the constant component flux in the I band is $F_\text{I}^\text{c}=10$~mJy, then it is unlikely that the variable component will create a brightness change in the observed wide range and exceeds $F_\text{I}^\text{c}$ is more than eight times.
On the other hand, the constant component is emission from a jet part in which there is motion along curved trajectories, as detected by VLBI data \citep{Rastorgueva09, Britzen09,Rani15, Lister16}.
Hence, the constant component may change over a long time interval, for example, due to its Doppler factor enhancement.
Then, in the observer's reference frame, the constant component flux increase with the spectral index $\alpha_\text{c}$ unchanged.
We plotted lines on the flux-flux diagrams corresponding to $\alpha_\text{c}=\text{const}$ (Fig~\ref{fig:A_const}).
Near $F_\text{I}=10$~mJy, small changes in $F_\text{I}^\text{c}$ lead to a strong one in $\alpha_\text{c}$.
We use the approximations of observed data, then we can't specify the exact $\alpha_\text{c}$ value, but it fulfills the condition $\alpha_\text{c}\gtrsim1.4$.
The point, corresponding to $F_\text{I}^\text{c}$ for the constant component minimum Doppler factor $\delta_\text{c}$, shifts to the right along the line for a constant $\alpha_\text{c}$ during the growth of $\delta_\text{c}$.
We use $F_\text{I}=10$, 15 and 20~mJy and $\alpha_\text{c}=1.5$ for further calculations.

\section{Obtaining a spectrum of the variable component from the observational data} \label{sec:avarobs}

\begin{figure}
\plotone{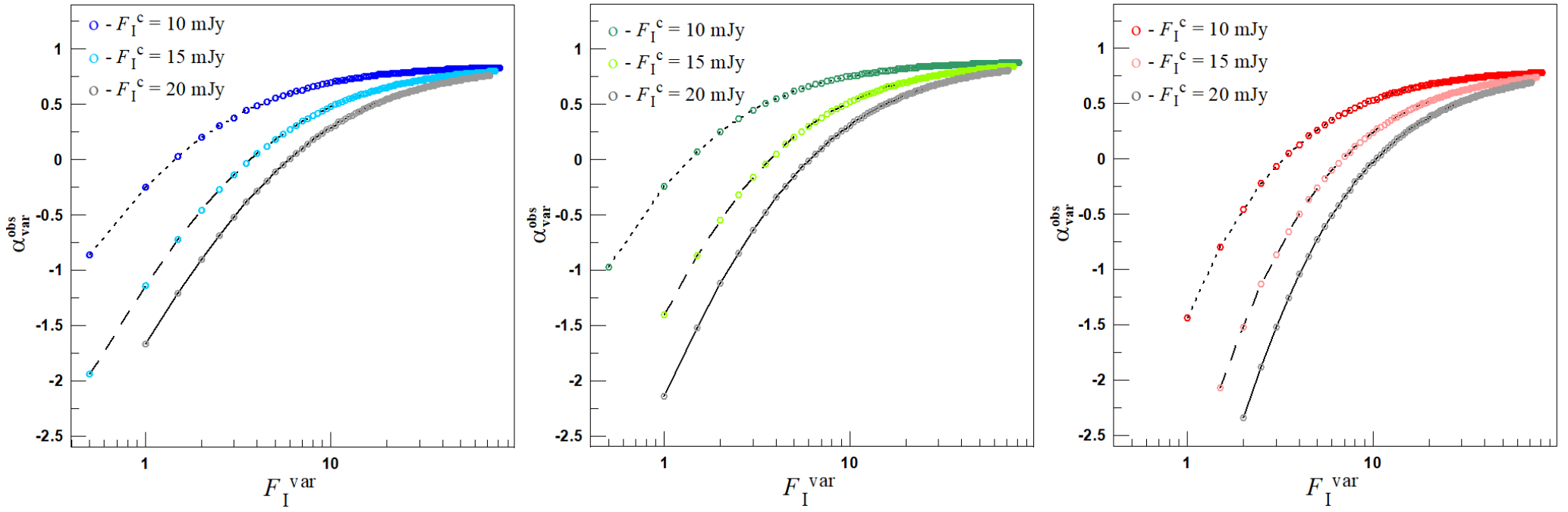}
\caption{Dependence of $\alpha^\text{obs}_\text{var}$} on the variable component flux in the I band for filter pairs B---I (left), V---I  (middle), R---I (right). The lines show the general trend of the points, which we plotted for different values of $F_\text{I}^\text{c}$. 
\label{fig:Alpha_obs_all}
\end{figure}

To calculate the spectral index of the variable component from the observational data, we assumed that each point on the observed line is the sum of the constant and variable component fluxes.
Then the variable component flux was determined by the formula:
\begin{equation} 
\label{eq:Fvar}
      	F_i^{\text{var}}=F_i^{\text{obs}}-F_i^{\text{c}},
\end{equation}
where $F_i^\text{obs}$ is the flux in some band $i$, which we define according to the expressions (\ref{eq:obs_line}).
We fixed $F_\text{I}^\text{c}$. 
There is a point on the observed line where $F_\text{I}=F_\text{I}^\text{c}$, then from the equation~(\ref{eq:Fvar}) we get that $F_\text{I}^\text{var}=0$.
Moving to the right with a step of 0.5~mJy along the observed line, we determine the observed spectral index of the variable component between the effective frequencies of the band I and $i$
\begin{equation} 
\label{eq:alpha_th1}
        \alpha_\text{var}^\text{obs}=-\frac{\lg\left(F_i^{\text{var}}/F_\text{I}^{\text{var}} \right)}{\lg\left(\nu_i/\nu_\text{I} \right)},
\end{equation}
where the index $i$ indicates that the value refers to the considered band $i$.
For low fluxes of $F_\text{I}^\text{var}$, $\alpha^\text{obs}_\text{var}$ takes an impossible small values.
With further $F_\text{I}^\text{var}$ increase, we reach a certain point, at which $\alpha^\text{obs}_\text{var}=-2.5$. This value corresponds to the radiation spectrum from an optically thick medium under synchrotron self-absorption.
From this moment, the values of $\alpha^\text{obs}_\text{var}$ for points on the observed line increase and are consistent with the synchrotron radiation theory.
We select these points to further analysis. 
Figure~\ref{fig:Alpha_obs_all} shows the dependency of $\alpha^\text{obs}_\text{var}$ on $F_\text{I}^\text{var}$.
Equation~(\ref{eq:alpha_th1}) defines a spectral index under the assumption of a power-law emission spectrum between the frequencies of $\nu_i$ and $\nu_\text{I}$. Then, the actual spectrum curvature is manifested in the difference of the spectral index determined between different frequencies.

\section{Case of a convex emission spectrum}
\subsection{One emission region} \label{sec:4.1}

\begin{figure}
\plotone{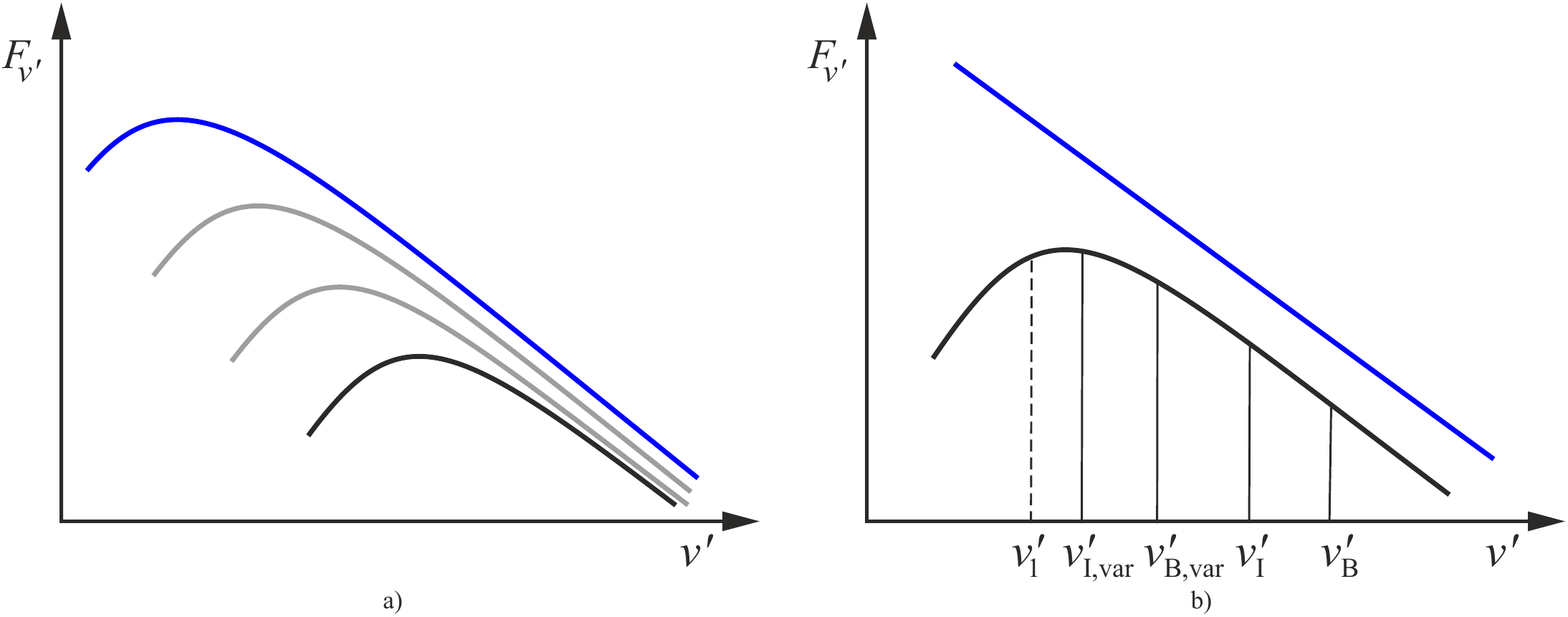}
\caption{The schemes of the observed blazar spectrum (left panel), composed of different synchrotron self-absorption emission regions, and the optical core spectrum (right panel) in the reference frame of comoving with relativistic jet plasma. 
On both panels, blue and black lines show the overall blazar emission and optical core spectra, respectively.
On the left panel, the gray curves display the spectra of distinct emission regions, for which the optical depth $\tau=1$ at different frequencies smaller than optical ones.
The radiation from the constant and variable components at the frequency range of $\nu^\prime_\text{I}-\nu^\prime_\text{B}$ and $\nu^\prime_\text{I,\,var}-\nu^\prime_\text{B,\,var}$, respectively, fall into the observed frequency range $\nu_\text{I}-\nu_\text{B}$.
$\nu^\prime_1$ is the frequency for which $\tau=1$.} 
\label{fig:spectr}
\end{figure}

Assuming the physical nature of the variable component, we obtain the theoretical dependence of the spectral index on the flux.
According to the \citet{Blandford} model, the farther away from the true jet base, the longer the wavelength for which the medium becomes transparent.
\citet{Pushkarev2012} confirmed that the VLBI cores observed at 15~GHz for several hundred sources correspond to this assumption.
This fact is consistent with the synchrotron self-absorption, which, as shown by \citet{Lobanov98}, is the dominant process of radiation absorption in jets.
Extrapolating this result to optical frequencies, we assume that all observed flux comes from the jet region, in which the medium becomes transparent to optical radiation (see Discussion for reasons to make this assumption).
Under a jet base, we mean a location where the extended on tens of parsecs and collimated ultrarelativistic outflow begins.
We emphasize that this region is approximately at a constant distance from the true jet base.
By analogy with the VLBI jet, we call this region the optical core.
Figure~\ref{fig:spectr} (left panel) shows a schematic spectrum of the jet parts with optical depth $\tau=1$ at several adjacent frequencies.
The total radiation of these regions gives the observed power-law spectrum at the optical-$\mu$m band, which flatters at mm-cm wavelengths.
If the optical core were resolved separately, the spectrum of these regions would have the form shown in Figure~\ref{fig:spectr} (right panel).

The constant component emission with the frequencies from $\nu_\text{I}^\prime$ to $\nu_\text{B}^\prime$ in the source reference frame falls into the fixed effective frequencies of the filters from $\nu_\text{I}$ to $\nu_\text{B}$.
We assume that the frequencies from $\nu_\text{I}^\prime$ to $\nu_\text{B}^\prime$ belong to the optically thin part of the spectrum.
Due to the jet curvature and non-radial motion detected for the S5~0716+714 parsec-scale jet \citep{Bach05, Britzen09, Rastorgueva09, Rastorgueva11, Rani15}, the jet region corresponding to the constant component might have different Doppler factors at different times.
We assume that for a rarely observed low state of the object with $F^\text{c}_\text{I}=10$~mJy, the Doppler factor of the constant component is $\delta_\text{c}=5$.
Increase of $\delta_\text{c}$ leads to enhancement of $F^\text{c}$.
For instance,  $\delta_\text{c}=6$ and 7 correspond to fluxes $F_\text{I}=15$ and 20~mJy, respectively.
Development of (magneto)hydrodynamic instabilities and turbulence is possible in the jet flow. 
Therefore, the emission region contains parts, which deviate from general motion direction and, hence, have Doppler factors $\delta_\text{var}$, which differs from $\delta_\text{c}$.
Emission from these parts having $\delta_\text{var}>\delta_\text{c}$ we consider as a variable component. 
At different times, the corresponding to constant emission component jet region has a different number of parts with the different $\delta_\text{var}$.
We assume that these parts produce the observed variability.
The higher Doppler factor $\delta_\text{var}$ leads to flux increase in the observer's reference frame, and radiation with frequencies $\nu^\prime_\text{I,\,var}-\nu^\prime_\text{B,\,var}$ falls into the observed frequency range $\nu_\text{I}-\nu_\text{B}$ (Fig.~\ref{fig:spectr}).
Figure~\ref{fig:spectr} shows that the spectrum between the frequencies $\nu^\prime_\text{I,\,var}-\nu^\prime_\text{B,\,var}$ is flatter than the spectrum of the constant component, which leads to the observed change in the color index.
Let's find out whether this interpretation applies to the observed dependence $\alpha_\text{var}^\text{obs}\left(F_\text{I}^\text{var}\right)$.

The spectral flux of the constant and variable components, taking into account synchrotron self-absorption, is \citep{Pachol}
\begin{equation}
    F_i=\delta^{2-5/2} Q^\prime  \nu^{5/2}_i\left(1+z\right)^{-3+5/2}\left\{1-\exp\left[-\left( \frac{\nu_i \left(1+z \right)}{\delta\cdot \nu^\prime_1} \right)^{-\alpha_\text{c}-5/2} \right] \right\},
    \label{eq:Fnu}
\end{equation}
with substitution the corresponding $\delta$ and $Q^\prime$.
In the Formula~(\ref{eq:Fnu}), $\nu^\prime_1$ is the frequency for which $\tau=1$, $\alpha_\text{c}$ is the constant component spectral index, which corresponds to the index of the optically thin part of the spectrum, $z$ is the object's redshift.
The coefficient $Q^\prime$ depends on the magnetic field, the particle number density, and the observed solid angle of the source.
We introduced the parameter characterizing the relative contribution of the variable component to the total radiation
\begin{equation}
    D=\frac{F_i^\text{var}}{F_i^\text{c}}\approx\frac{d\Omega^\prime_\text{var}}{d\Omega^\prime_\text{c}},
    \label{eq:D1}
\end{equation}
where $d\Omega^\prime_\text{var}$ and $d\Omega^\prime_\text{c}$ are the observed solid angles of the variable and constant components in the comoving reference frame.
We have a single radiating region, and a variable component is some part of it, having a higher Doppler factor during some time. 
So, in the simplest assumption about uniform emission region, the magnetic field, $B$, and particle number density, $n_e$, are the same for the constant and variable components. 
Then, the flux ratio of the variable to the constant component is the ratio of $d\Omega^\prime_\text{var}$ and $d\Omega^\prime_\text{c}$.
Although some fluctuations are possible, in our view, the change of the magnetic field configuration mainly influences the ratio because the transverse to the line of sight magnetic field component enters in Equation~(\ref{eq:Fnu}) as part of $Q^\prime$ and $\nu^\prime_1$.
For $n_e=const$, the changing $\nu^\prime_1$ by order of magnitude requires a 20-fold change in the magnetic field component perpendicular to the line of sight, $B_\bot$. 
Such decreasing or increasing in $B_\bot$ gives a multiplier of 1.5 or 0.2  on the right side of the formula~(\ref{eq:D1}). To realize such extreme $B_\bot$ changes, a strictly ordered magnetic field directed at a small angle to the line of sight is necessary, which is unlikely.

For the fixed value of $F_i^\text{c}$, we calculate $F_i^\text{var}$, taking into account (\ref{eq:Fnu}) and (\ref{eq:D1}).
For this, we were changing the values of $D$ from 0.01 to 1 in increments of 0.01, $\delta_\text{var}$ from $\delta_\text{c}$ to 40 in increments of 0.1, $\nu^\prime_1$ from $10^{13}$ to $10^{14}$~Hz in increments of $0.5\cdot10^{13}$~Hz.
The upper limit for $\delta_\text{var}$ was chosen as the maximum possible value~\citep{Butuzova18a, Liodakis15}.
We selected the interval of $\nu^\prime_1$ based on our assumption about the emitting region of the constant component, accounting that the upper bound of which is such that the radiation at the observed frequencies approximately corresponds to the optically thin spectrum part.
If the total flux in each considered band falls into $\pm2\sigma$ from the observed line  (where $\sigma=0.65$~mJy is the maximum standard deviation out of the three values obtained from the flux-flux diagram for three filter pairs), then the theoretical spectral index of the variable component $\alpha_\text{var}^\text{th}$ between the effective frequencies of the filters $i$ and I was calculated using a formula similar to the expression~(\ref{eq:alpha_th1}).
If the condition $\alpha_\text{var}^\text{obs}-0.02\leq\alpha_\text{var}^\text{th}\leq\alpha_\text{var}^\text{obs}+0.02$ was fulfilled, then we select a point characterized by a set of fixed values $F_\text{I}^\text{var}$, $F_i^\text{var}$, $\nu^\prime_1$, $\delta_\text{var}$, $\alpha_\text{var}^\text{th}$, and $D$ for further analysis.

\begin{figure}
\plotone{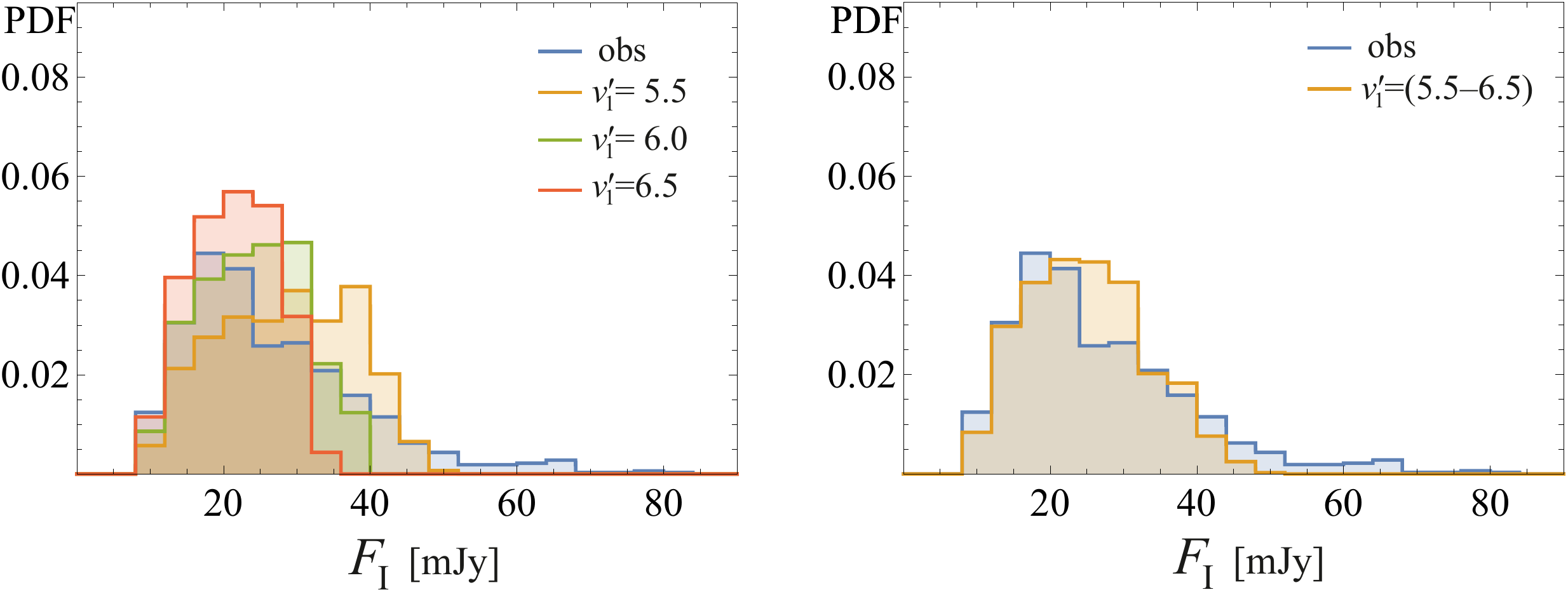}
\caption{Probability density functions of the observed and theoretical fluxes: a) for the fixed $\nu^\prime_1$; b) for the interval of $\nu^\prime_1=(5.5 - 6.5)\cdot10^{13}$~Hz. Both histograms show the functions for the observed flux in blue.} 
\label{fig:v1_all}
\end{figure}

In the framework of our assumptions, the radiating region locates at a constant distance from the true jet base.
Various parts of the jet flow pass through this region.
If the jet has a global configuration of the magnetic field, then the change of Doppler factor of this region leads to the change of perpendicular to the line of sight magnetic field component \citep{Lyutikov05}.
For a constant magnetic field strength, this fact causes a fluctuation of $\nu^\prime_1$ within some small interval \citep{Pachol}.
To select the values of $\nu^\prime_1$ for further analysis, we compared the observed distribution of $F_\text{I}^\text{obs}$ with the theoretical $F_\text{I}^\text{th}$ one.
Under this, for each value of $\nu^\prime_1$, we combine the theoretical points obtained for the considered pairs B---I, V---I, R---I for the fixed $\nu^\prime_1$.
We selected $\nu^\prime_1$ in the range $(1-9.5)\cdot 10^{13}$~Hz.
For small $\nu^\prime_1$, there is a uniform distribution over the entire observed range, up to the maximum flux.
With increasing frequency, a peak appears in the distribution of $F_\text{I}^\text{th}$ at high values of $\nu^\prime_1$, which further gradually shifts to low values of $\nu^\prime_1$.
There are almost no theoretical points for $\nu^\prime_1$, which escape beyond the considered interval.
The distribution of $F_\text{I}^\text{th}$ for frequencies from $5.5\cdot 10^{13}$ to $6.5\cdot 10^{13}$~Hz agrees well with the observed one (Fig.~\ref{fig:v1_all}).
Therefore, we selected theoretical points for which $\nu^\prime_1$ lies in the range of $(5.5 - 6.5)\cdot10^{13}$~Hz for further analysis.
Note that the observed data with the values $F_\text{I}>50$~mJy are very rare and make up less than $5\%$ of the total point number.
Within the framework of our assumptions, theoretical points with $F_\text{I}>50$~mJy exist at a lower $\nu^\prime_1$ than the selected interval.

\begin{figure}
\plotone{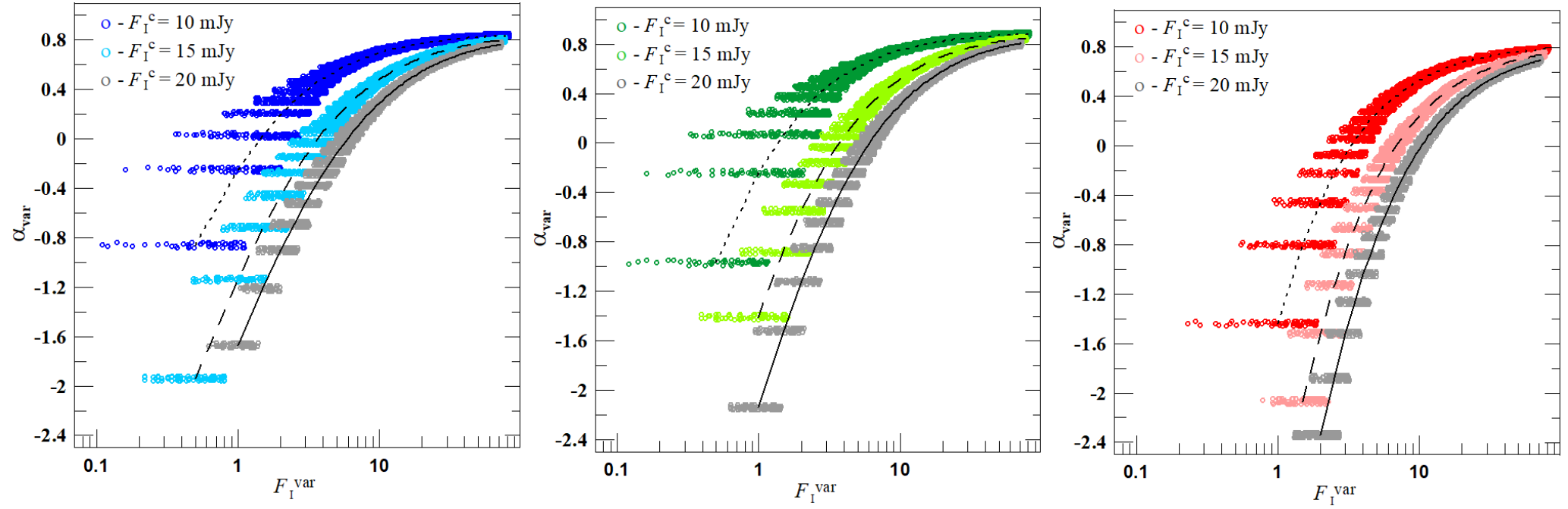}
\caption{The dependence of $\alpha_\text{var}$ on $F_\text{I}^\text{var}$ for $\nu^\prime_1=(5.5 - 6.5)\cdot10^{13}$~Hz for a pair of bands B---I (left), V---I (middle), R---I (right). The symbols show the simulated points. The lines correspond to the dependency for $\alpha_\text{var}^\text{obs}$.} 
\label{fig:Av(Fv)_5}
\end{figure}

Simulation results $\alpha_\text{var}^\text{obs}\left(F_\text{I}^\text{var}\right)$ are shown in the figure~\ref{fig:Av(Fv)_5}.
For $F_\text{I}^\text{var}<1$~mJy, a slight increase in the flux gives a rapid growth of $\alpha_\text{var}$.
Although there are no theoretical points for $F_\text{I}\gtrsim50$~mJy, we described $\approx95\%$ of the observational data.

\begin{figure}
\plotone{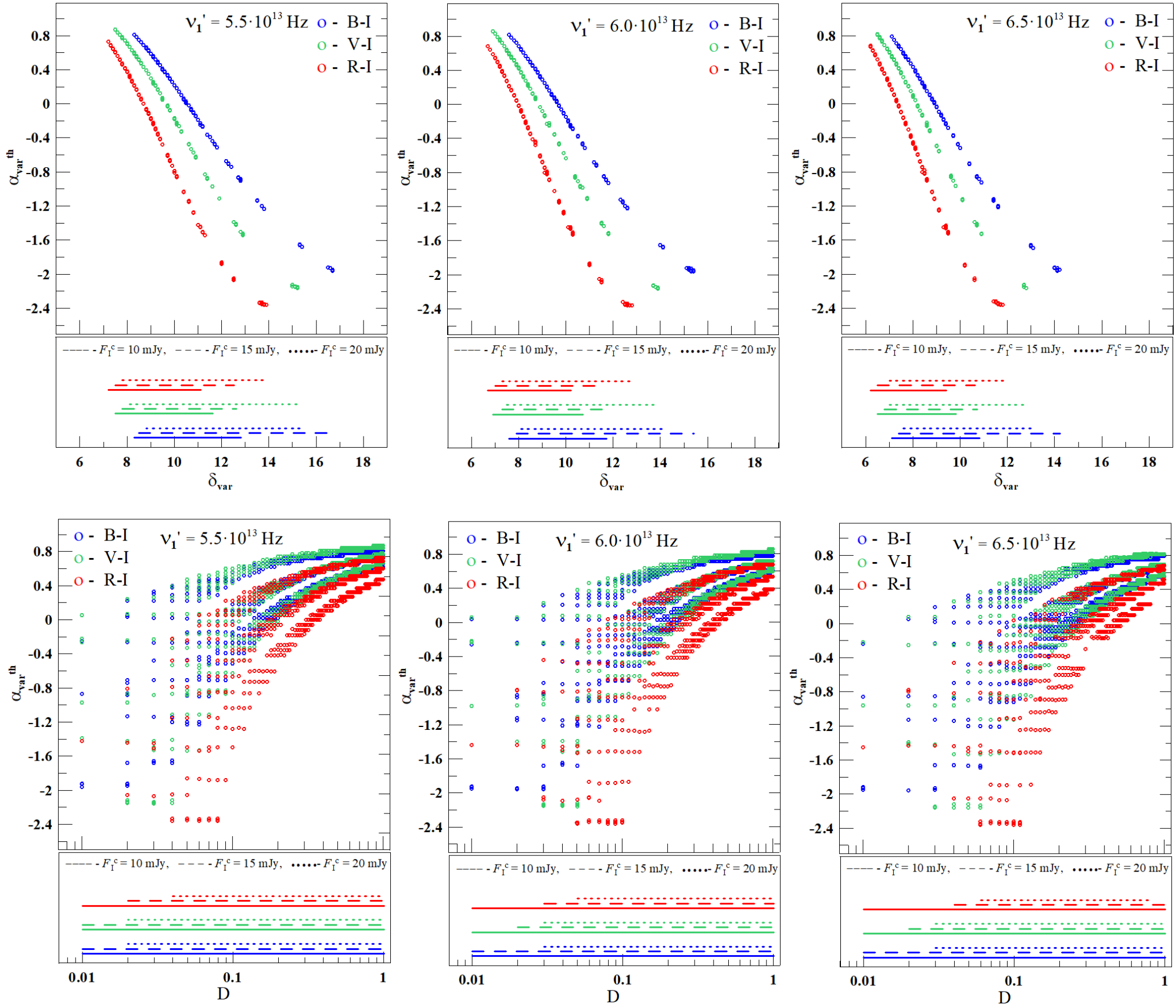}
\caption{
The dependencies of $\alpha_\text{var}^\text{th}$ on $\delta_\text{var}$ (top panels) and $D$ (bottom panels) for different  $\nu^\prime_1$.
On each plot, the intervals of possible values of $\delta_\text{var}$ and $D$ are given for different $F_\text{I}^\text{c}$.} 
\label{fig:Av(d_D)}
\end{figure}

Let us consider the dependence of $\alpha_\text{var}^\text{th}$ on other used parameters: $\delta_\text{var}$ and $D$ (Fig.~\ref{fig:Av(d_D)}). 
On the plot of $\alpha_\text{var}^\text{th}\left(\delta_\text{var}\right)$, the points lie along one line for all $F_\text{I}^\text{c}$ and the fixed pair of bands.
For $F_\text{I}^\text{c}=10$~mJy, the range of $\delta_\text{var}$ is the smallest.
With an increase in $\nu^\prime_1$, the interval of doppler variation shifts slightly towards low values, while always $\delta_\text{var}>\delta_\text{c}$.
Within the assumption on the radiating region, that occurs if the variable component doesn't move along some directions.
This fact imposes additional conditions on the jet matter motion.
In addition, for a fixed $\delta_\text{var}$, $\alpha_\text{var}^\text{th}$'s for the filter pairs under consideration have different values. 
It indicates that the spectrum of the variable component is always convex, which is naturally expected for $\delta_\text{var}>\delta_\text{c}$.

Points for each filter pair form the three oblong curved areas on the distribution  $\alpha_\text{var}^\text{th}$ on $D$ (Fig.~\ref{fig:Av(d_D)}, bottom panel).
Each point group is associated with the defined value of $F_\text{I}^\text{c}$.
With the increase of $F_\text{I}^\text{c}$, $\alpha_\text{var}^\text{th}$ of the theoretical points and the range of possible values of $D$ decrease, and the low boundary of the latter shifts slightly towards higher $D$.
Since $\alpha_\text{var}^\text{th}<\alpha_\text{c}$, radiation comes from the region where the absorption is significant.
For the fixed $F_\text{I}^\text{c}$, the ranges for $\delta_\text{var}$ and $D$ are determined by the intersection of the corresponding intervals for all filter pairs.
Thus, the variable component can cover a volume from a small part to the almost entire optical emitting region. At the same time, the Doppler factor changes from $\approx7$ to 14.

\subsection{Two emission regions} \label{sec:4.2}

Let us consider two optical emitting regions, namely, the optical core and the extended jet. 
We took the compact region with the curved spectrum (the optical core) as the variable component.
The jet radiation has a power-law spectrum, because its medium is optically thin.
Often for interpretation of the observed properties \citep[e.g.][]{Bach05, Rastorgueva2011, Rani15, Lister13, Lister16, Butuzova18a,Butuzova18b,Butuzova2020jet}, the jet is assumed to be curved, which implies different Doppler factors for its various parts.
Since we consider the jet as a whole and it is more extended than the optical core, the possible variability of the total radiation flux of the jet caused by geometric effects occurs on longer time scales and with a smaller amplitude of the brightness change.
Therefore, we can consider the jet as the constant component.

Based on the arguments about the minimum flux of the constant component (Section~\ref{subsec:2.2}), we accept $\alpha_\text{c}\approx1.5$ and $F_\text{I}^\text{c}=10$~mJy.
We do not consider large values of $F_\text{I}^\text{c}$ since we believe that the average jet Doppler factor does not change significantly.
From observations, it is impossible to determine the relative brightness of the core and the jet in the source reference frame.
VLBI observations S5~0716+714 show a significant domination of the radiation of the VLBI core in the total flux from the source \citep{Britzen09,Rastorgueva09, Rani15}.
Then there would be $F_\text{I}^\text{var}\gtrsim10$~mJy, and therefore the total flux from the object would be greater than observed one.
From the observations, the values of $F_\text{I}^\text{var}<10$~mJy are possible and correspond to the situation when the radiation flux of the optical core in the observer's reference frame is less than the jet flux.
Assuming the influence of only geometric effects, the flux from the initially faint jet can prevail over the core flux if the Doppler factor of the optical core $\delta_\text{var}$ is significantly less than the jet's Doppler factor $\delta_\text{c}$.
In calculations, we take $\delta_\text{c}=15$ and the minimum Doppler factor of the variable component $\delta_\text{var, min}=5$.
On the other hand, as shown by~\citet{Kovalev20, Petrov17}, blazars can have bright optical jets that outshine the radiation of the optical core.
Here we also consider this case, assuming $\delta_\text{c}=5$.
It leads us to consider cases with the faint and bright jet.

\begin{figure}
\plotone{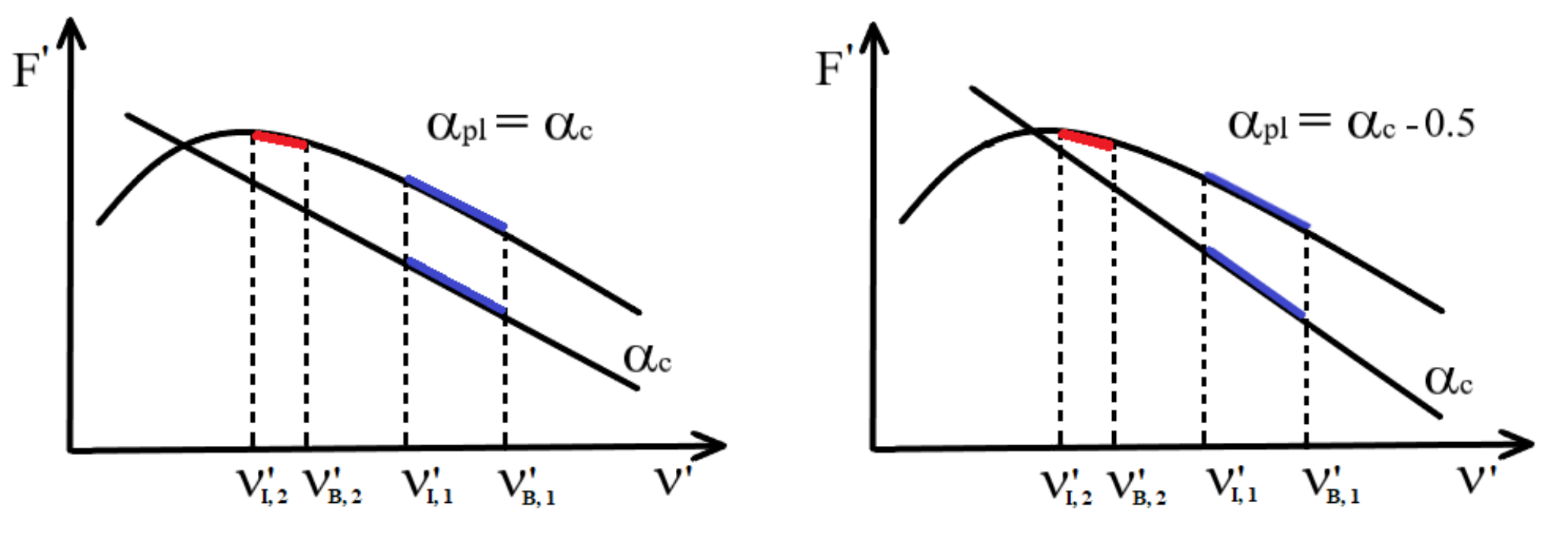}
\caption{Spectrum schemes of the constant and variable components.
The left panel illustrates the case in which electrons do not undergo significant radiation losses propagating from the core to the jet.
The right panel refers to the case in which emitting jet electrons have undergone significant radiation losses.
When $\delta_\text{var, min}=\delta_\text{c}$, the observed frequencies $\nu_\text{I}-\nu_\text{B}$ correspond to the radiation with $\nu_{\text{I}, 1}^\prime-\nu_{\text{B}, 1}^\prime$ in the source reference frame (highlighted by blue).
With an increase in $\delta_\text{var}$, the radiation of the variable component at lower frequencies of $\nu_{\text{I}, 2}^\prime-\nu_{\text{B}, 2}^\prime$ in the source reference frame  (highlighted by red) falls into the observed frequency range.    
} 
\label{fig:v2_all}
\end{figure}

In both cases, we do not know the spectral index of both the jet radiation $\alpha_\text{c}$ and the optically thin part of the variable component spectrum $\alpha_\text{pl}$.
It is natural to assume that $\alpha_\text{pl}=\alpha_\text{c}$ (Fig.~\ref{fig:v2_all}, left panel).
On the other hand, electrons, propagating from the optical core into the jet, can undergo significant energy losses through radiation, which will lead to a steepening of their energy spectrum.
Then $\alpha_\text{pl}=\alpha_\text{c}-0.5$ (Fig.~\ref{fig:v2_all}, right panel).

In our assumptions, the flux variability is due to a change in the core Doppler factor $\delta_\text{var}$.
Increasing $\delta_\text{var}$ will shift the curved spectrum of the variable component to the lower frequencies in the source reference frame, which leads to a decrease in $\alpha_\text{var}$ calculated by the formula (\ref{eq:alpha_th1}) between the two considered frequencies, that is, to the BWB-trend (Fig.~\ref{fig:v2_all}). 
For a numerical description of this scenario, we introduce a parameter that characterizes the relative contributions of the radiation of the variable and constant components to the total flux
\begin{equation} 
\label{eq:D2}
      	D_F=\frac{F_\text{var, min}^\prime}{F_\text{c}^\prime},
\end{equation}
where the index ``min'' denotes the flux at $\delta_\text{var, min}=5$.
The variable component curved spectrum means that the values of $D_F$ differ for $F_\text{var, min}^\prime$ and $F_\text{c}^\prime$ in different filters, therefore for clarity in the definition of \text{$D_F$}, we used the fluxes at the frequency $\nu_\text{I}$.
Assuming that $F_\text{I}^\text{c}=10$~mJy are associated with $\delta_\text{c}=15$ and 5 for faint and bright jet, respectively, we found $F_\text{c}^\prime$.
The values of $F_\text{var, min}^\prime$ were determined using the expression
\begin{equation}
    F_\text{var, min}^\prime=\delta_\text{var, min}^{-5/2}\cdot Q_\text{var}^\prime  \,\nu^{5/2}\left(1+z\right)^{5/2}\left\{1-\exp\left[-\left( \frac{\nu \left(1+z \right)}{\delta_\text{var, min}\cdot \nu^\prime_1} \right)^{-\alpha_\text{pl}-5/2} \right] \right\}.
    \label{eq:Fvar0}
\end{equation}
For the fixed $F_\text{c}^\prime$, we substituted (\ref{eq:Fvar0}) into (\ref{eq:D2}) and, varying $D_F$ from 0.1 to 60 in 0.1 increments, found $Q_\text{var}^\prime$. 
Then $Q_\text{var}^\prime$'s we used to calculate $F_\text{var}$ by the expression (\ref{eq:Fnu}) with the substitution of the corresponding parameters.
We changed the values of $\delta_\text{var}$ from $\delta_\text{var, min}$ to 40 in increments of 0.1, $\nu^\prime_1$ from $10^{13}$ to $10^{14}$~Hz in increments of $0.5\cdot10^{13}$~Hz.
The upper limit for $\delta_\text{var}$ was the maximum possible value \citep{Butuzova18a, Liodakis15}.
We chose the interval of $\nu^\prime_1$ based on our assumptions about the variable component, and the upper bound is such that $\nu^\prime_1<\nu_\text{I}^\prime$ under $\delta_\text{var, min}$.
Theoretical points, which represent a set of parameters ($F_\text{I}^\text{var}$, $F_i^\text{var}$, $\nu^\prime_1$, $\delta_\text{var}$, $\alpha_\text{var}^\text{th}$, and $D_F$), were selected according to the algorithm described in Section~\ref{sec:4.1}.
The simulation results for various $\alpha_\text{pl}$ are in the following subsections.

\subsubsection{$\alpha_\text{pl}=\alpha_\text{c}$} \label{sec:4.2.1}

\begin{figure}
\plotone{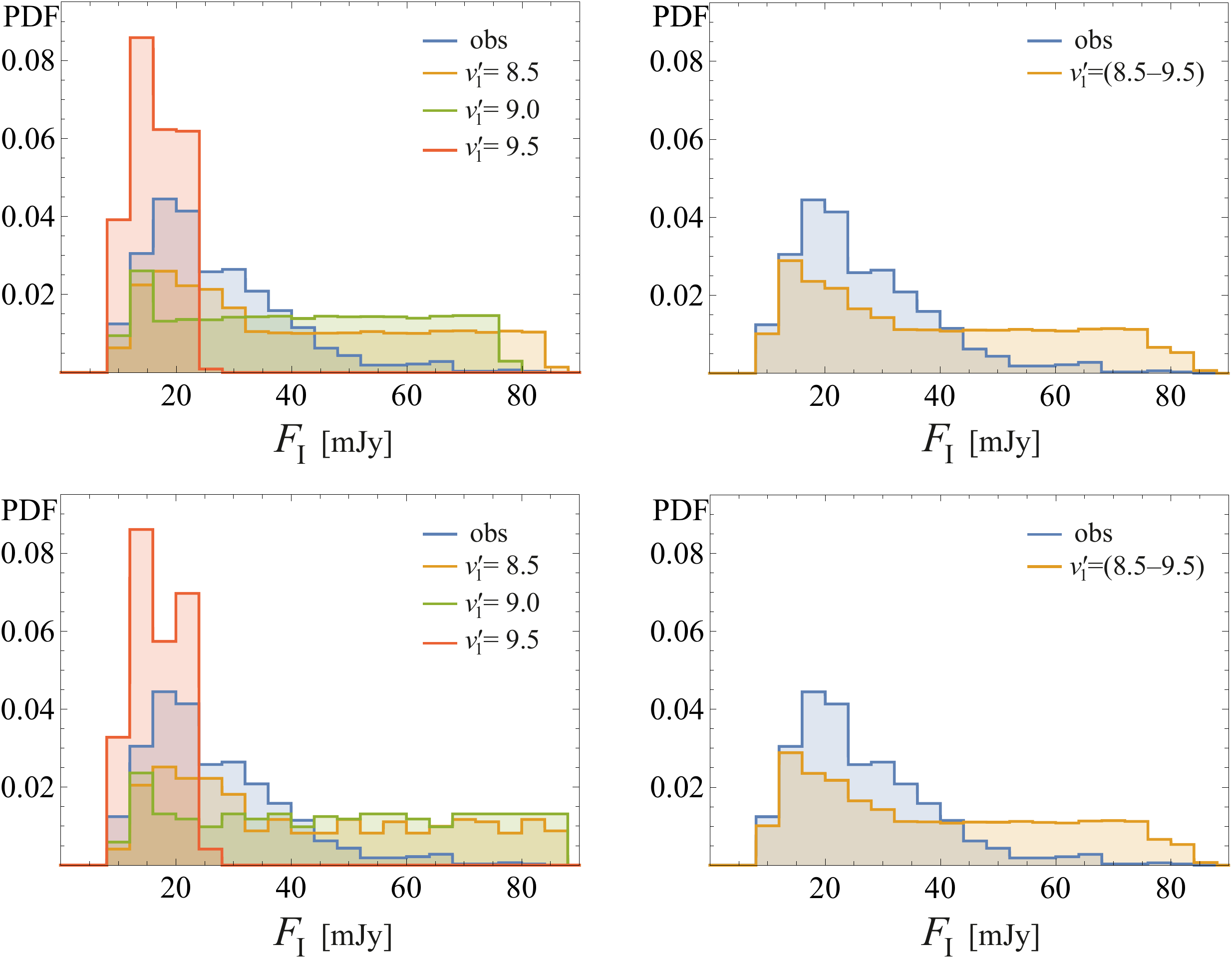}
\caption{
The probability density functions of observed and theoretical fluxes for the bright (upper panel) and faint (lower panel) jets under $\alpha_\text{pl}=\alpha_\text{c}$. The distributions for fixed $\nu^\prime_1$ are at left panels, for the interval $\nu^\prime_1=(8.5 - 9.5)\cdot10^{13}$~Hz --- at right ones. On all histograms, the distribution of the observed flux is showed in blue.
} 
\label{fig:v2.1_Nu}
\end{figure}

First, we considered the case when the electrons have not significant energy losses as they traveled from the optical core to the jet.
Combining the theoretical points obtained for the considered filter pairs at the fixed $\nu^\prime_1$ for both bright and faint jet states, we found that for $\nu^\prime_1<~8.5\cdot~10^{13}$~Hz, the distribution of the theoretical flux is flat. As $\nu^\prime_1$ increases, a peak on the total flux of 15---20~mJy appears and becomes sharper (Fig.~\ref{fig:v2.1_Nu}). The distribution for frequencies in the range of $(8.5-9.5)~\cdot~10^{13}$~Hz has the best agreement with the observed.

\begin{figure}
\plotone{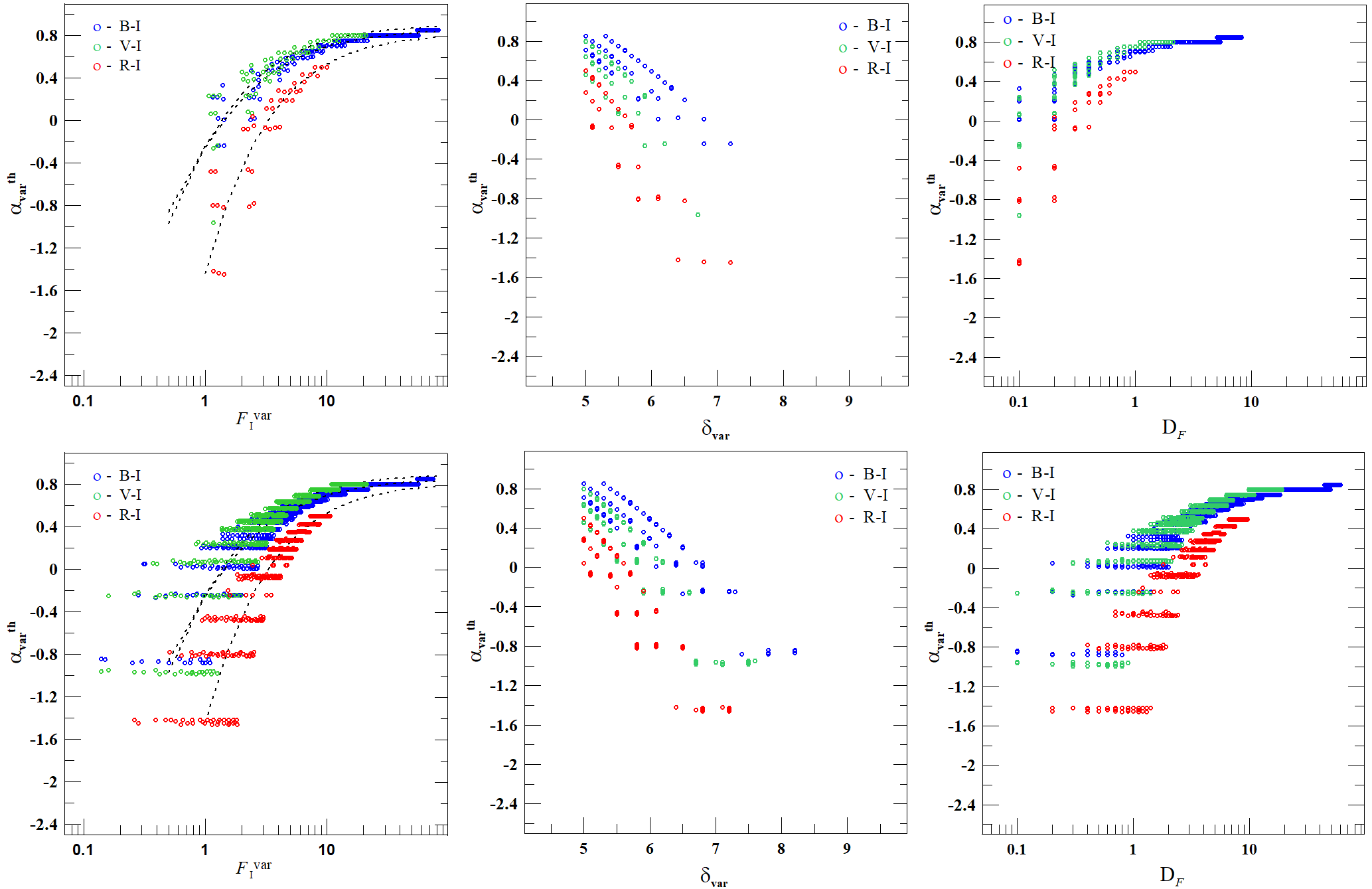}
\caption{
The dependencies $\alpha_\text{var}^\text{th}$ on $F_\text{I}^\text{var}$ (left panels), $\delta_\text{var}$ (middle panels) and $D_F$ (right panels). The top and bottom panels relate to the cases of the bright and faint jet. Points on the plots refer to the selected frequency range $\nu^\prime_1=(8.5-9.5)\cdot10^{13}$~Hz and $\alpha_\text{pl}=\alpha_\text{c}$.
} 
\label{fig:v2.1_all}
\end{figure}

Figure~\ref{fig:v2.1_all} shows the simulation results for the selected frequency range $\nu^\prime_1$.
For the faint jet, there are points at $F_\text{I}^\text{var}<1$~mJy and, on the whole, noticeably more points than for the bright one (Fig.~\ref{fig:v2.1_all}, left panels). 
In both cases, the values of $F_\text{I}>30$ and 20~mJy don't realize for the filter pairs V---I and R---I, respectively.
The smaller values of $\nu^\prime_1\approx(7.5-8.0)\cdot10^{13}$~Hz are needed to achieve such fluxes.

There are three groups of points on the plots $\alpha_\text{var}^\text{th}\left(\delta_\text{var}\right)$ (Fig.~\ref{fig:v2.1_all}, middle panels), corresponding to $\nu^\prime_1=8.5$, $9.0$ and $9.5\cdot10^{13}$~Hz, for each considered filter pair.
As $\nu^\prime_1$ increases, the range of $\delta_\text{var}$ decreases and shifts to lower $\delta_\text{var}$ values.
In cases of the faint and bright jet, the resulting range of possible values of $\delta_\text{var}$ is approximately from 5 to 7.5.
For any $\delta_\text{var}$, the radiation comes from a medium with an optical thickness of $\approx1$.
Values of $\delta_\text{var}\approx5$ realized for $D_F>1$ consistent with the $\delta_\text{var, min}$ for the cases of both bright and faint optical jets.
Performed modeling does not give an advantage to any of the considered assumptions about the jet Doppler factor.

\subsubsection{$\alpha_\text{pl}=\alpha_\text{c}-0.5$} \label{sec:4.2.2}

\begin{figure}
\plotone{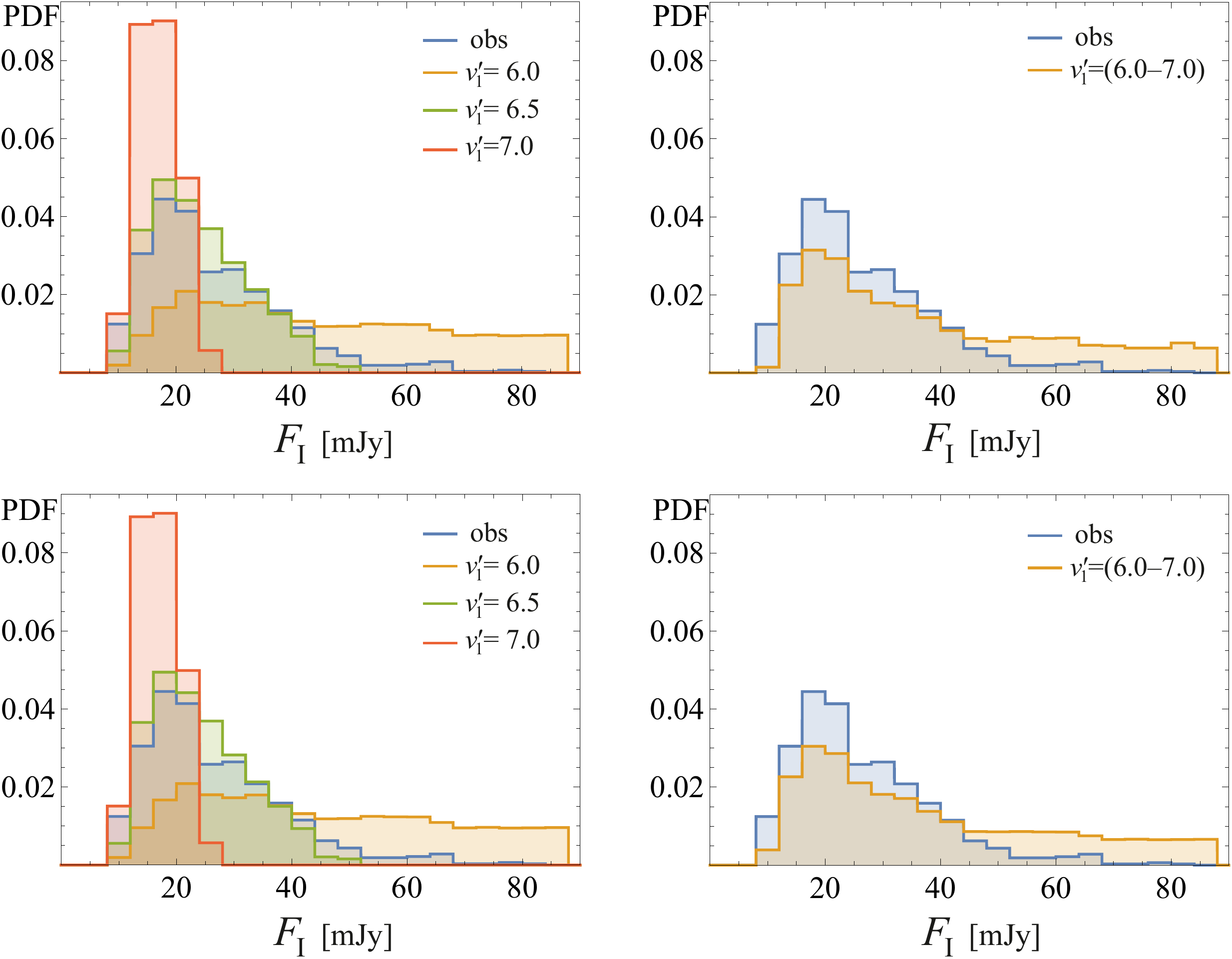}
\caption{The probability density functions of the observed and theoretical fluxes for the bright (upper panel) and faint (lower panel) jets under $\alpha_\text{pl}=\alpha_\text{c}-0.5$. 
 The distributions for fixed $\nu^\prime_1$ are at left panels, for the interval $\nu^\prime_1=(6.0-7.0)\cdot10^{13}$~Hz --- at right ones. On all histograms, the observed flux distribution is showed in blue.} 
\label{fig:v2.2_Nu}
\end{figure}

Here we considered the case of significant electron energy losses through radiation when electrons travel from the optical core to the jet.
Similar to the previous cases (Sec.~\ref{sec:4.1} and \ref{sec:4.2.1}), we combine the theoretical points for the considered filter pairs at the fixed $\nu^\prime_1$ and choose the frequency range that best describes the observed distribution.
Figure~\ref{fig:v2.2_Nu} showed histograms of the probability density function for the theoretical and observed fluxes.
At $\nu^\prime_1<6\cdot10^{13}$~Hz, the distribution of theoretical fluxes is almost flat, and the weakly manifested maximum slowly shifts towards large values of $F_\text{I}$ with a decrease of $\nu^\prime_1$.
With the increase of $\nu^\prime_1$, the peak is on the total flux values of 15---20~mJy. 
At the same time, the interval of $F_\text{I}$ changes decreases and shifts to the lower values.
Good agreement between the observed and theoretical flux distributions is at the frequency $6.5\cdot10^{13}$~Hz.
Corresponding modeled points for filter pairs of B$-$I, V$-$I, R$-$I describe approximately $95\%$, $55\%$, and $70\%$ of observed data, respectively, excepting high flux values. As we mentioned above, $\nu^\prime_1$ can change due to changes of emitting region Doppler factor or the transverse to the line of sight magnetic field component. 
So, to describe high fluxes and agreement between the observed and modeled points, we used the frequency range $(6-7)\cdot10^{13}$~Hz. 

\begin{figure}
\plotone{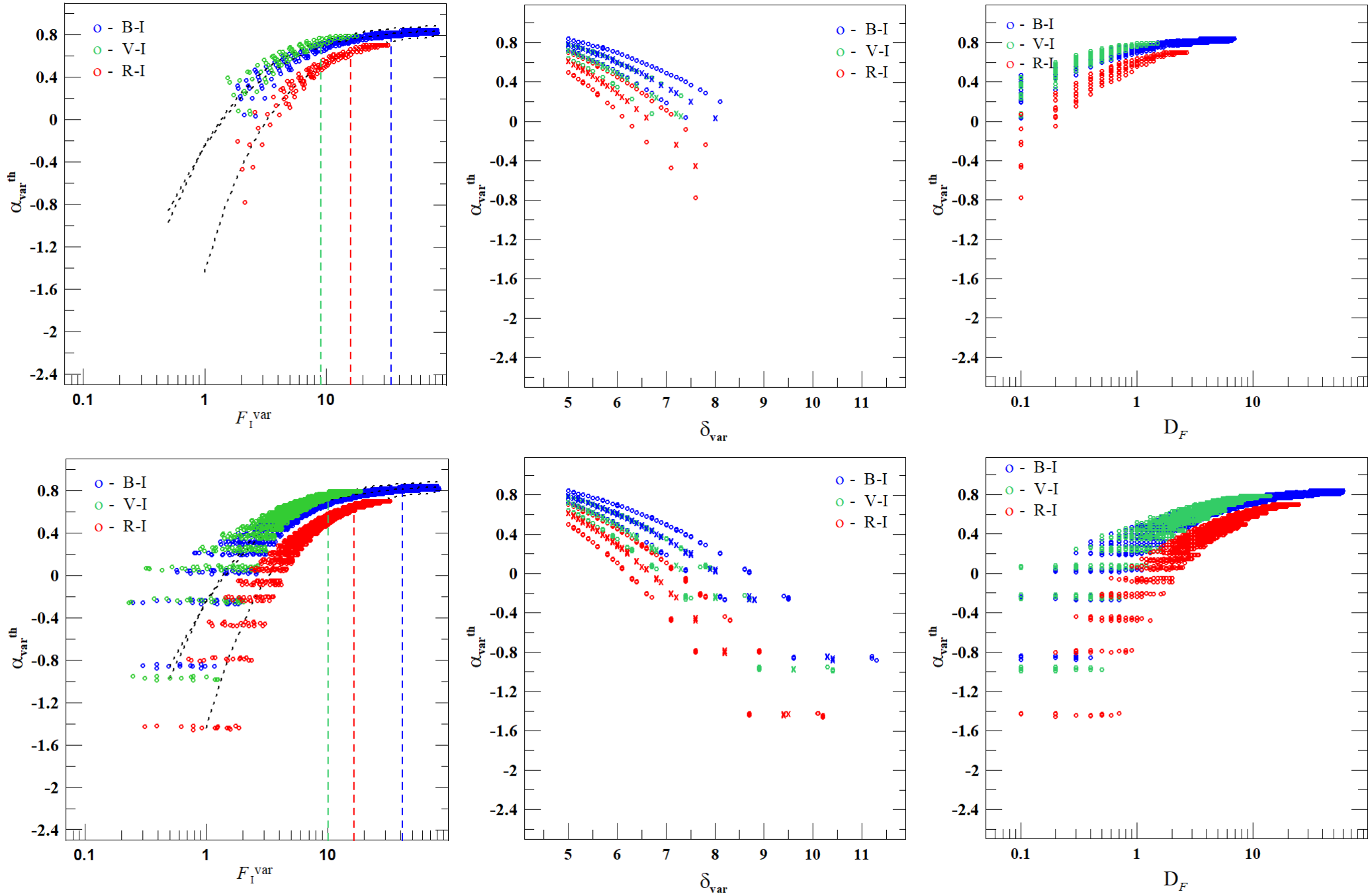}
\caption{The dependencies of $\alpha_\text{var}^\text{th}$ on $F_\text{I}^\text{var}$ (left panels), $\delta_\text{var}$ (middle panels), and $D_F$ (right panels) are presented for cases of the bright and faint jet on the upper and lower panels, respectively. Points on the plots refer to the selected frequency range $\nu^\prime_1=(6.0-7.0)\cdot 10^{13}$~Hz and $\alpha_\text{pl}=\alpha_\text{c}-0.5$.
Points of simulations for the frequency $\nu^\prime_1=6.5\cdot10^{13}$~Hz have maximum flux denoted for the corresponding filter pairs by colored vertical lines at the left panels. Crosses show these points at the middle panels. At the right panels, the points locate close to the center in the corresponding oblong areas.}

\label{fig:v2.2_all}
\end{figure}

The simulated dependence's of $\alpha_\text{var}$ on $F_\text{I}^\text{var}$, $\delta_\text{var}$, and $D_F$ for fixed $\nu^\prime_1$ in the selected range (Fig.~\ref{fig:v2.2_all}) almost do not differ from the cases considered in Section~\ref{sec:4.2.1}: (i) there are significantly more theoretical points for the case of the faint jet; (ii) there are no points for $F_\text{I}^\text{var}<1$~mJy for the bright jet; (iii) in both cases $F_\text{I}>20$~mJy for filter pairs V---I and R---I occur at $\nu^\prime_1<6.0\cdot10^{13}$~Hz.
If $\nu^\prime_1=6.5\cdot10^{13}$~Hz is only considered, then the results slightly change. Namely, at the left panel of Fig.~\ref{fig:v2.2_all}, points with high flux will be absent, and distributions of $\alpha_\text{var}(\delta_\text{var})$  and $\alpha_\text{var}(D_F)$ will become narrower.

From the dependence of $\alpha_\text{var}^\text{th}$ on $\delta_\text{var}$ (Fig.~\ref{fig:v2.2_all}), it follows that for small $\delta_\text{var}$, the medium of the variable component is already practically transparent to radiation at the observed frequencies ($\alpha_\text{var}^\text{th}\approx0.8\approx\alpha_\text{pl}$), in contrast to results in Section~\ref{sec:4.2.1}.
The significant difference between the cases $\alpha_\text{pl}=\alpha_\text{c}$ and $\alpha_\text{pl}=\alpha_\text{c}-0.5$ is the different $\nu^\prime_1$.
Assuming the equal size of the radiating region, different $\nu^\prime_1$'s correspond to different magnetic field strengths.
Therefore, the magnetic field estimation can indicate the most probable case.

\section{Magnetic field}

The assumption of synchrotron self-absorption at the known $\nu^\prime_1$ allows us to estimate the component of the magnetic field perpendicular to the line of sight $B_\bot$~\citep{Slish}.
At the same time, the angular size of our source is unknown.
We suppose the optical emission region locates near the true jet base, and its size is comparable to the gravitational radius of the black hole.
\citet{Woo02, Pian05, Sbarrato12} estimated the black hole masses in blazars, which range from $10^8$ to $5\cdot10^8\:\text{M}_\odot$, and for S5~0716+714, the estimated mass of the black hole is $M_\text{BH}=10^{7.87}\:\text{M}_\odot$ \citep{Zhang2012}.
Therefore, we used for further calculations $M_\text{BH}=10^8$ and $5\cdot10^8\:\text{M}_\odot$.

\begin{figure}
  \centering
\includegraphics[scale=0.7]{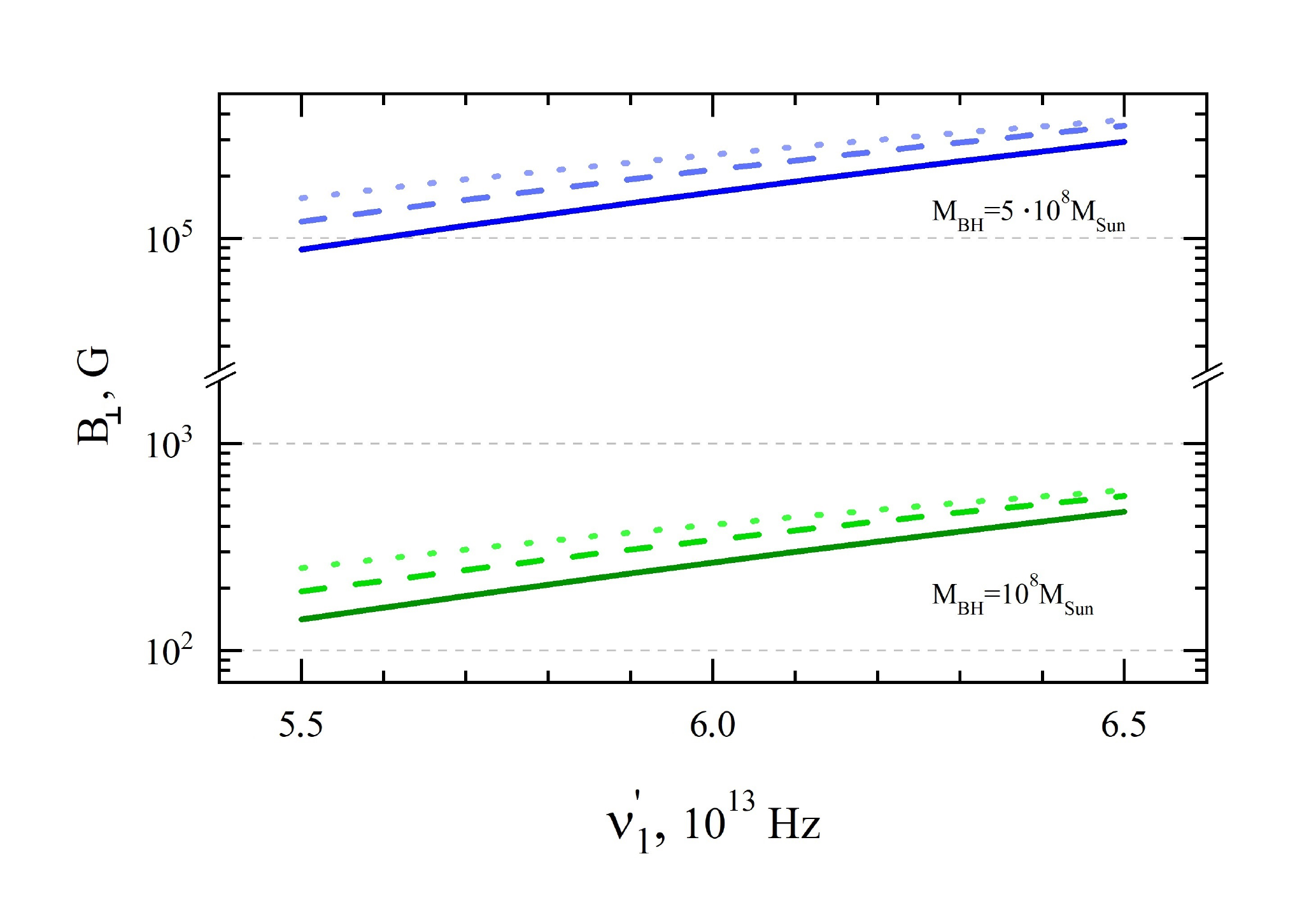}
\caption{$B_\bot(\nu^\prime_1)$ for the case of the single radiating region, having the size of one gravitational radius of the black hole with mass of $10^8\:\text{M}_\odot$ (green lines) and $5\cdot10^8\:\text{M}_\odot$ (blue lines). Solid, dashed, and dotted lines show the dependencies for $F_\text{I}^\text{c}=10$, 15, and 20~mJy, respectively.
} 
\label{fig:v1_mag}
\end{figure}

The magnetic filed component perpendicular to the line of sight in the source reference frame is \citep{Butuzova21}
\begin{equation} 
\label{eq:magfield}
       B_\bot=\left(\frac{\pi}{4}\frac{c_5(\alpha_\text{pl})}{c_6(\alpha_\text{pl})} \right)^2 \left\{ 1-\exp\left[-\left(\frac{\nu_i(1+z)}{\delta\cdot\nu^\prime_1}\right)^{-\alpha_\text{pl}-5/2}\right] \right\}^2(2c_1)^{-5}\, \frac{R_g^{\prime\,4}}{D_L^4} F_i^{-2}\: \nu_i^5 (1+z)^{-1} \delta^{-3}, 
\end{equation}
where $R_g^\prime$ is the emission region size in the source reference frame, $D_L$ is the luminosity distance of the object, functions $c_5(\alpha_\text{pl})$ and $c_6(\alpha_\text{pl})$  are tabulated by \citet{Pachol}.
Using Formula~(\ref{eq:magfield}) for the case of the single emission region, we substituted the obtained points with $F_\text{I}^\text{c}$, $\delta_\text{c}$, and $\alpha_\text{c}=1.5$ at the selected $\nu^\prime_1$ (Section~\ref{sec:4.1}) instead of $F_i$, $\delta$, and $\alpha_\text{pl}$, respectively. The calculated magnetic field depending on $\nu^\prime_1$ for different emission region sizes is displayed in Figure~\ref{fig:v1_mag}. 

\begin{figure}
\plotone{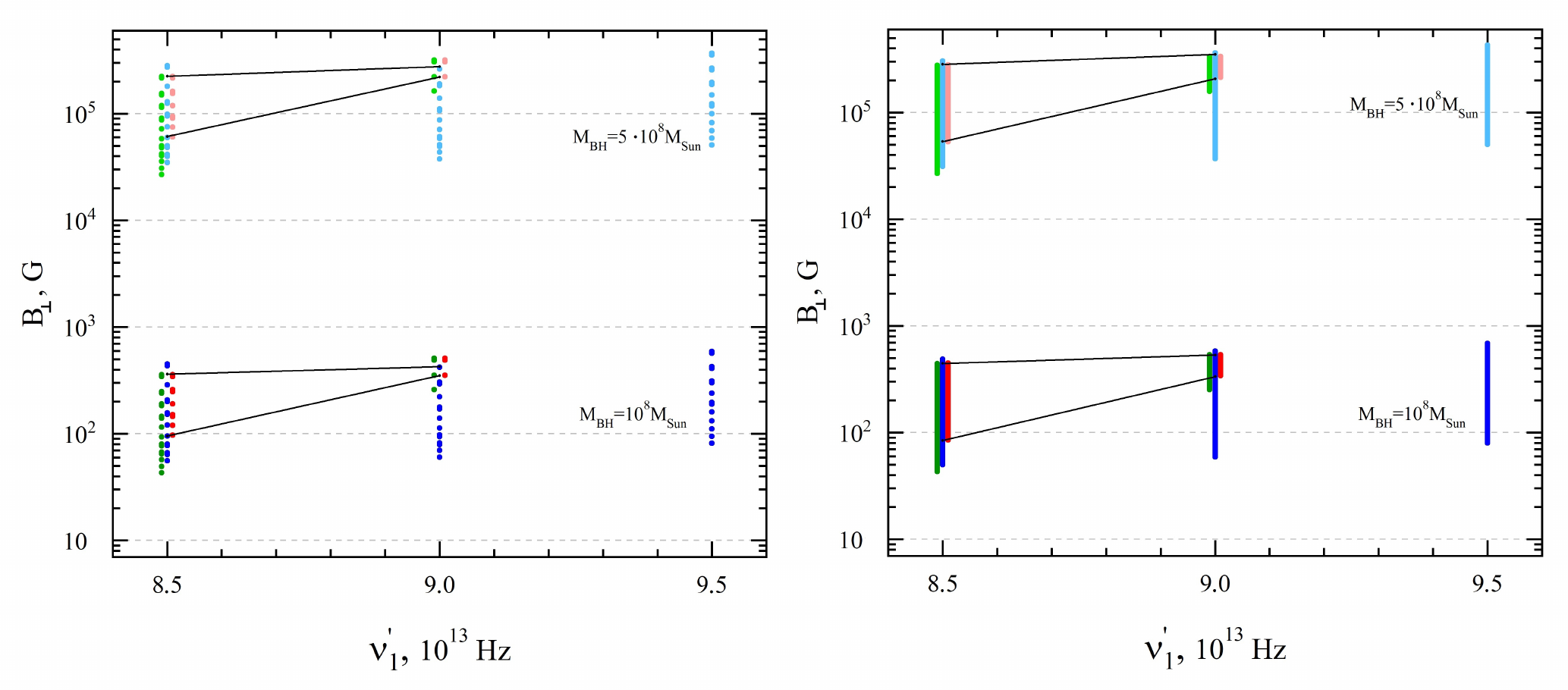}
\caption{The dependencies of $B_\bot$ on $\nu^\prime_1$ for bright (left panel) and faint (right panel) jet under $\alpha_\text{pl}=\alpha_\text{c}$. The blue, green, and red colors indicate the values of $B_\bot$ obtained for the filter pairs B---I, V---I, and R---I, respectively. The lines connect the boundaries of the intervals of the $B_\bot$ values at the fixed $\nu^\prime_1$.} 
\label{fig:v2.1_mag}
\end{figure}

For the case of two radiating regions, the synchrotron self-absorption acts in the variable component.
We use obtained in Sections~\ref{sec:4.2.1} and \ref{sec:4.2.2} points with $F_\text{I}^\text{var}=5$---15~mJy because the maximum distribution of total observed flux (Fig.~\ref{fig:v2.1_Nu}) occurs at $F_\text{I}=15$---20~mJy.
As Figure~\ref{fig:v2.1_mag} shows, $B_\bot$ takes a high value for the case of $\alpha_\text{pl}=\alpha_\text{c}$ (Section~\ref{sec:4.2.1}). 
In contrast, the magnetic field is acceptable for both the bright and faint jet in the case of $\alpha_\text{pl}=\alpha_\text{c}-0.5$ (Section~\ref{sec:4.2.2}). Especially, $B_\bot$ well agrees with other independent estimates for $M_\text{BH}\approx10^8\:\text{M}_\odot$ \citep{mag_field1, mag_field2}.

\begin{figure}
\plotone{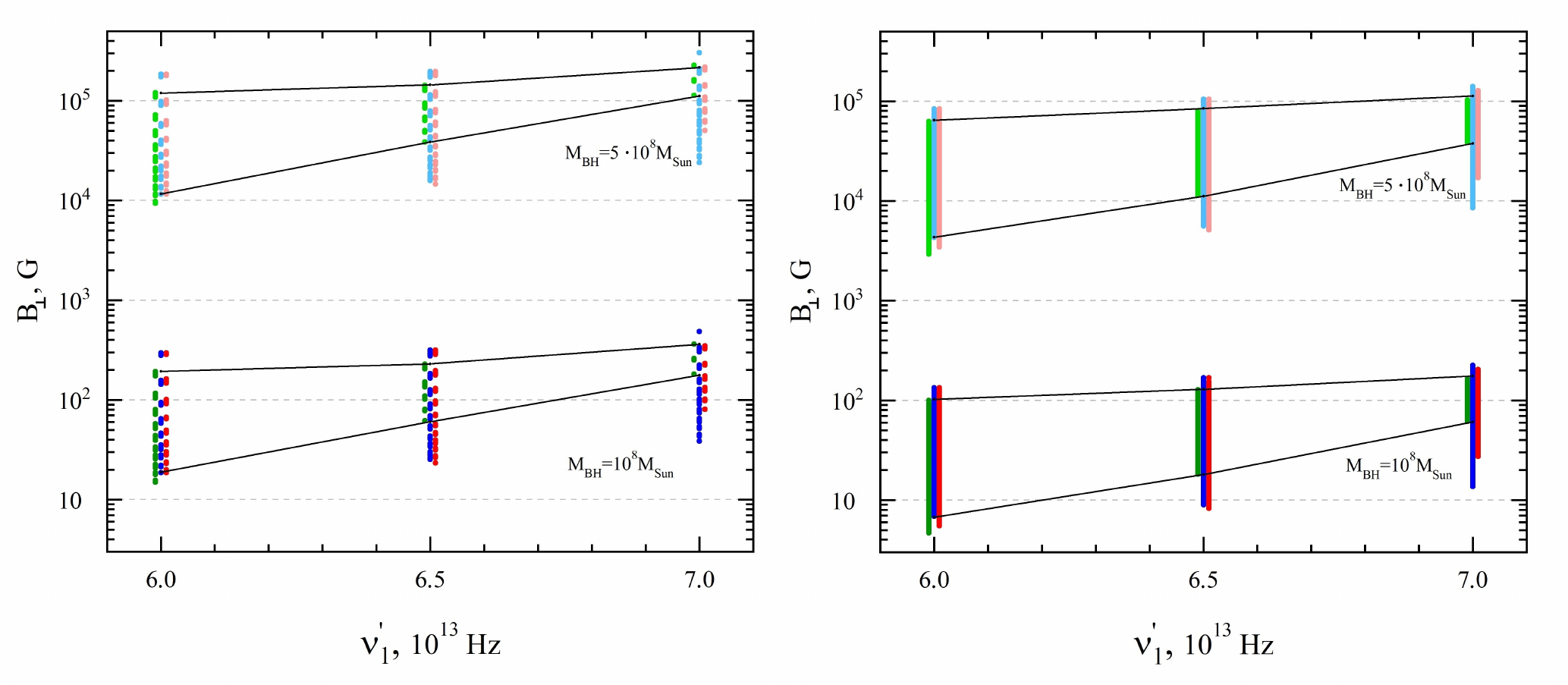}
\caption{The dependencies of $B_\bot$ on $\nu^\prime_1$ for bright (left panel) and faint (right panel) jet under $\alpha_\text{pl}=\alpha_\text{c}-0.5$. The blue, green, and red colors indicate the values of $B_\bot$ obtained for the filter pairs B---I, V---I, and R---I, respectively. The lines connect the boundaries of the intervals of the $B_\bot$ values at the fixed $\nu^\prime_1$.} 
\label{fig:v2.2_mag}
\end{figure}

\begin{figure}
    \centering
    \includegraphics[scale=0.8]{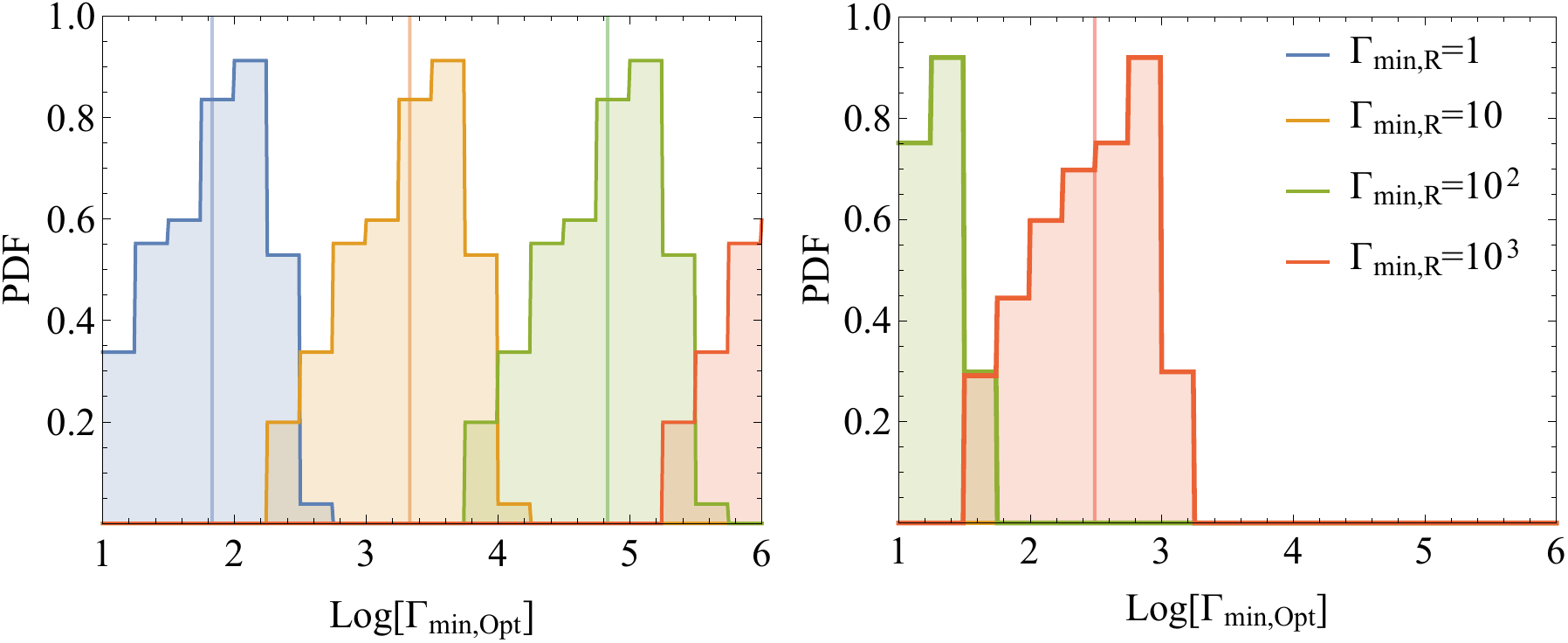} 
    \caption{
    Probability density functions for $\Gamma_\text{min, Opt}$'s, which are needed to fulfill $n_{e, \text{Opt}}=n_{e, \text{Opt}}^\text{extr}$ for different values of $\Gamma_\text{min, R}$. For plots, we used the emitting region size of the gravitational radius of the black hole with the mass of $10^8 M_\odot$ (left panel) and $5 \cdot 10^8 M_\odot$ (right panel). Colored vertical lines mark the median of the corresponding distribution.
    }
    \label{fig:edens}
\end{figure}

The final analysis that would confirm the action of synchrotron self-absorption in the optically emitting region is an estimation of electron density $n_{e,\text{Opt}}$.
To date, there are no estimates of the absolute values of $n_{e,\text{Opt}}$, and $n_{e,\text{Opt}}$ depends on the choice of the lower boundary of the power-law electron energy distribution. Hence, we compared the extrapolation of the electron density in the 15~GHz core to the optically emitting region with the density calculated using the found parameters. Electron density in the 15~GHz core is
\begin{equation}
    n_{e, \text{R}}=\frac{N_{0, \text{R}}}{\gamma_\text{R}-1} \left(m c^2\right)^{1-\gamma_\text{R}} \Gamma_\text{min, R}^{1-\gamma_\text{R}},
    \label{eq:ne}
\end{equation}
where $\Gamma_\text{min, R}$ is the minimum Lorentz factor of power-law electron energy distribution in the 15~GHz core, $\gamma_\text{R}$ is the spectral index of electron distribution. The constant $N_{0, \text{R}}$ is \citep{Pachol} 
\begin{equation}
    N_{0,\text{R}}=\left(\frac{\nu^\prime_1}{2 c_1} \right)^{(\gamma_\text{R}+4)/2} (s\, c_6)^{-1} B_\bot^{-(\gamma_\text{R}+2)/2},
    \label{eq:N0}
\end{equation}
where $s$ is the size of emitting region in the source reference frame. 
We used Formula~(\ref{eq:N0}) with parameters of 15~GHz VLBI core, namely, $B_\bot=0.07$~G, $\delta=10$ \citep{Pushkarev2012}, $\alpha=1.5$. 
To obtain $s$, we estimated the angular size of the 15~GHz core of $s_\text{mas}=0.038$~mas at the epoch of 2006 May 24 \citep[used by][]{Pushkarev2012}, modeling the core by the circular Gaussian component.
The $\nu^\prime_1$ we found assuming that the spectral maximum at $\nu_\text{m}=15$~GHz. 
Using the relation of the distance from the true jet base with the region of $\tau_\nu=1$ as a function of frequency and the quadratic decline of the electron density \citep{Lobanov98}, the electron density extrapolated to the optical core is
\begin{equation}
    n_{e, \text{Opt}}^{\text{extr}}=n_{e, \text{R}}\left( \frac{\nu_\text{Opt}}{\nu_\text{R}}\right)^2 \approx 4 \cdot 10^8 \, n_{e, \text{R}}.
    \label{eq:ne_extr}
\end{equation}
On the other hand, we calculated the electron density in the optical core $n_{e, \text{Opt}}$ using the obtained $\nu^\prime_1$, $B$, $\delta$, assuming that the emitting region size is equal to the gravitational radius of the black hole.
We performed these calculations using Formulae (\ref{eq:ne}) and (\ref{eq:N0}) with substitution of the values corresponding to the most probable assumption for radiating regions (Section~\ref{sec:4.2.2}).
For several values of $\Gamma_\text{min, R}$, we calculated the minimum Lorentz factor of the power-law electron energy distribution in the optical core $\Gamma_\text{min, Opt}$ based on $n_{e, \text{Opt}}=n_{e, \text{Opt}}^\text{extr}$ and obtained a set of minimum Lorentz factors of the power-law electron energy distribution in the optical core ($\Gamma_\text{min, Opt}$), which probability density distribution is displayed in Fig.~\ref{fig:edens}. 
It can be seen that for the case of $M_\text{BH}=10^8 M_\odot$, the $\Gamma_\text{min, Opt}$'s are higher than the corresponding $\Gamma_\text{min, R}$. 
It can be explained by that the electrons traveling from optical to 15~GHz VLBI core lose their energy through synchrotron radiation. 
On the other hand, for the case of $M_\text{BH}=5 \cdot 10^8 M_\odot$, the $\Gamma_\text{min, Opt}$'s are somewhat smaller than the corresponding $\Gamma_\text{min, R}$. 
Therefore, it is possible that for some mass of the black hole between $(1-5)\cdot M_\odot$, the condition $\Gamma_\text{min, Opt}=\Gamma_\text{min, R}$ will be fulfilled. 
Therefore, the obtained $\Gamma_\text{min, Opt}$'s seem adequate, especially under the assumption of $M_\text{BH}=10^8 M_\odot$, which indicates realistic parameters of the jet region in which synchrotron self-absorption of optical radiation acts.

\section{Cases of non-power-law electron spectrum}

\begin{figure}
\centering
\includegraphics[scale=0.8]{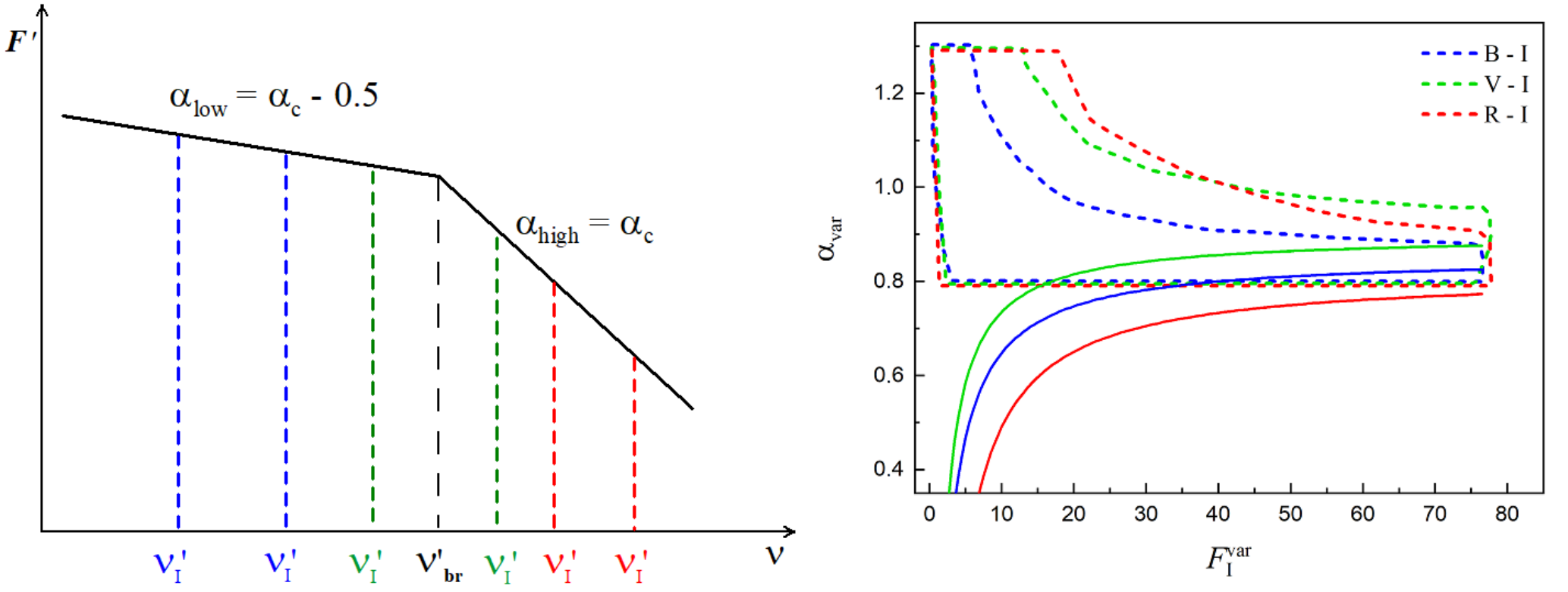}
\caption{Left panel: the spectrum of the radiating region with a break at a frequency of $\nu^\prime_\text{br}$ in the reference frame comoving with} the relativistic jet. The higher the Doppler factor, the lower frequencies fall into the observed range. Therefore, the spectral index varies from $\alpha_\text{c}$ to $\alpha_\text{c}-0.5$. Right panel: the theoretical and observed distributions of $\alpha_\text{var}$ on $F_\text{I}^\text{var}$. The dashed lines indicate the areas corresponding to the obtained theoretical points. The solid lines show the dependence according to the observational data. The blue, green, and red colors indicate the filter pairs B---I, V---I, and R---I, respectively. 
\label{fig:sbreak}
\end{figure}

Jet ultrarelativistic electrons lose energy through radiation, the rate of which is proportional to the square of the particle energy.
Therefore, a spectral break appears in the initially power-law electron energy distribution at its upper boundary \citep{Pachol}. 
Over time, the Lorentz factor of electrons $\Gamma_\text{br}$, at which the break occurs, decreases.
Electrons with $\Gamma_\text{br}$ generate radiation at the frequency $\nu^\prime_\text{br}$, at which the break is in the radiation spectrum.
There is a relation for spectral indices before and after the break  $\alpha_\text{low}=\alpha_\text{high}-0.5$.
Suppose that at a minimum value of $\delta$, photons with a frequency of $\nu^\prime>\nu^\prime_\text{br}$ in the source reference frame fall into the observed frequencies (Fig.~\ref{fig:sbreak}, left panel).
Then $\alpha_\text{obs}=\alpha_\text{high}$.
The higher the Doppler factor of radiating region, the lower photon frequency $\nu^\prime$ falls into the observed frequencies.
For $\nu^\prime_\text{I}<\nu^\prime_\text{br}<\nu^\prime_\text{B}$, the values of the observed spectral index are $\alpha_\text{low}<\alpha_\text{obs} <\alpha_\text{high}$.
If $\nu^\prime_\text{B}<\nu^\prime_\text{br}$, $\alpha_\text{bs}= \alpha_\text{low}$ and it does not change with a further increase of $\delta$.
Let's find out if this scenario can reproduce the observed change in $\alpha_\text{var}$ depending on $F^\text{var}$.

We considered the whole radiating region, having a minimum Doppler factor $\delta_\text{c}$, as the constant component.
The variable one is a part of the radiating region that deviates from the general trajectory closer to the observer's direction and, as a result, has a higher Doppler factor $\delta_\text{var}\geqslant\delta_\text{c}$.
A point of $F_\text{I}=10$~mJy on the observed line corresponds to $\alpha\approx1.4-1.5$.
If this point is to take as the constant component, then the minimum value of $\alpha_\text{var}$ lies from 0.9 to 1, which exceeds all possible values of $\alpha_\text{var}^\text{obs}$, found in Section~\ref{sec:avarobs}.
There must be $\alpha_\text{c}\approx1.3$ for an intersection of $\alpha_\text{var}^\text{th}$ and $\alpha_\text{var}^\text{obs}$.
For $F_\text{I}=10$~mJy, the point with $\alpha_\text{c}\approx1.3$ lies above the observed line plotted for the filter pair I---V within the range of the value spread.
For the same logic of calculations of $F^\text{var}$ and $\alpha_\text{var}$ through this paper, for further analysis, we took $\delta_\text{c}=6$ and $F_\text{I}^\text{c}=15$~mJy, at which the point with $\alpha_\text{c}=1.3$ lies slightly below the observed lines defined for the considered filter pairs.
It does not affect the result qualitatively.

For modeling, we used $\nu^\prime_\text{br}$ from $4\cdot10^{13}$ to $6\cdot10^{13}$~Hz in increments of $0.2\cdot10^{13}$~Hz and $\delta_\text{var}$ from 5 to 50 in increments of 0.1. The parameter $D$, characterizing the ratio of the variable and constant component sizes, varied from 0.01 to 1 in increments of 0.01.
Figure~\ref{fig:sbreak} (right panel) shows the regions in which all the obtained theoretical points lie.
The agreement to the observed dependence $\alpha_\text{var}$ on $F_\text{I}^\text{var}$ occurs at large fluxes for all filter pairs.
The maximum of the observed flux distribution corresponds to 15---25~mJy, hence, the considered assumption cannot describe most points.
This result indicates that this case is unlikely over significant difficulties to interpret the observational data.
To produce the observed long-term dependence on the flux-flux diagram, a few broken power-law electron energy distributions with different spectral indices and ``fine-tuning'' $\delta_\text{var}$ for each electron population are necessary.
Note that we considered a simple case of a discrete break of the synchrotron power-law spectrum, but the obtained conclusion remains right for a curved power-law electron spectrum at high energies, which has been considered, e.g., by \cite{Tanihata04}.

\begin{figure}
\centering
\includegraphics[scale=0.7]{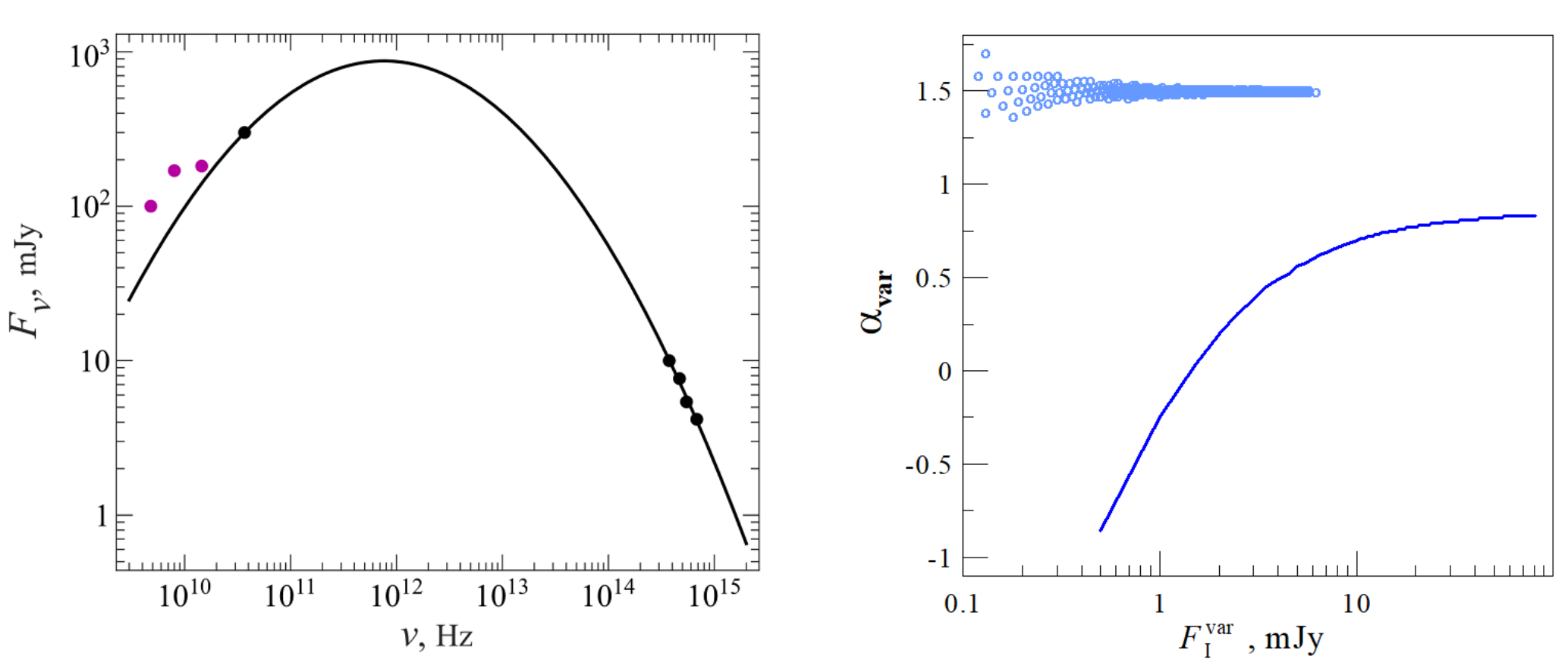}
\caption{
Left panel: radio and optical fluxes of the constant component in the blazar S5~0716+714 radiation and its approximation by log-parabolic function through only black circle (solid line). Purple points indicate minimum flux at frequencies of 4.8, 8, and 14.5~GHz, which are possibly influenced by the opacity effect and kiloparsec-scale structure radiation.
Right panel: theoretical (points) and observed (line) dependencies of variability component spectral index, determined between B- and I-filters. For other filter pairs, the dependencies are similar. 
} 
\label{fig:logpar}
\end{figure}

Another possibility for producing a curved photon spectrum is the acceleration process, in which the probability for a particle to be accelerated is higher when its energy is lower \citep{Massaro04a}. 
Such spectrum was applied to the explanation of high-energy hump in the spectral energy distribution of a few blazars \citep[see, e.g., ][]{Massaro04a,Massaro04b, Nieppola06, Tramacere07, Donnarumma09}. 
To obtain parameters of log-parabolic synchrotron spectrum we used the flux of optical constant component ($F_\text{B}$, $F_\text{V}$, $F_\text{R}$ were found by Equation~(\ref{eq:obs_line}) for $F_\text{I}=10$~mJy) and the minimum flux in the radio band at a frequency of 37~GHz (Fig.~\ref{fig:logpar}, left panel).
At this frequency and lower ones, there are observations over several decades \citep{Bychkova15}. We did not use lower frequency to avoid the probable opacity effect and contribution to the total flux from the extended radio structure of the blazar. 
We perform the analogous calculation of $\alpha_\text{var}^\text{th}$, varying $D$ from 0.01 to 1 in increments 0.01 and $\delta_\text{var}$ from 5 to 50 in increments of 0.1. There is no correspondence with the observational data (Fig.~\ref{fig:logpar}, right panel).
The fitting spectrum containing all shown in Fig.~\ref{fig:logpar} radio data points has a lesser bend, hence, it results in smaller spectral index changes under the variability caused by geometric effects.

\section{Discussion}

Although the object shows different color behavior under flux variability at various periods, the trend of the color change with brightness on scales of tens of years turns out to be surprisingly steady (from 02.2002 to 06.2019 by our data).
The data of \citet{Raiteri03} for the period from 11.1994 to 01.2002 are in good agreement with the obtained dependence on flux-flux diagrams (Fig.~\ref{fig:FF_RV_all}).
It indicates a general global process that leads to variability for several decades.
On shorter time scales, from intra-day to several months and years, the color index behavior with flux variability slightly differs from the long-term trend.
It can occur due to fluctuations of the bulk process or/and by the influence of another, less significant variability mechanism.

Under the standard assumption of power-law spectra, geometric effects do not change the object color index.
However, physical processes change the object spectrum.
If both mechanisms, characterized by the different behavior of the color index with brightness, make a comparable contribution to the variability, then there would be no well-marked dependence in the flux-flux diagram, which contradicts the observations.
Our observational data indicates that the color becomes bluer when the object is brighter (BWB-trend).
\citet{MarscherGear85} interpreted the BWB-trend by the physical process leading to variability.
But before analyzing jet physics, it is necessary to estimate how significant the contribution of geometric effects to variability, because they indeed operate in the object for the following reason.
All observed radiation from blazars comes mainly from its jet.
The non-radial motion of jet features \citep[see, e.g.][]{Kim2020,Britzen09,Homan01} and changes in the inner jet position angle for many AGN \citep{Lister13}, including S5~0716+714, were detected from VLBI observations.
It indicates that the jet is nonlinear on parsec-scale and may have a helical shape.
Hence, the angle of the velocity vector to the line of sight changes, and, consequently, the Doppler factor changes.
\citet{Britzen09} interpreted the long-term radio variability of S5~0716+714 by geometric effects. \citet{Rani13} concluded that the modulation of long-term optical variability of S5~0716+714 by geometrical effects is most probably.

Geometric effects can explain the BWB-trend if the electron or radiation spectrum is not power-law.
We obtained that the broken power-law and log-parabolic spectra cannot reproduce the observed change in the color index on the long-term scale of variability.
At the moment, synchrotron self-absorption for higher frequencies is not reliably confirmed. 
However, it can indirectly manifest in time delays of variability at lower frequencies compared to variability at higher frequencies, as \citet{Agarwal17} shown in the radio range for frequencies from 4.8 to 36.8~GHz.
Note, the jet helical shape reduces correlation \citep{Rani14,Butuzova18b}.
\citet{Rani13} detected a time delay of $\approx65^\text{d}$ between the fluxes in the V band and at the frequency of 230~GHz.
The flux variability of the VLBI core at millimeter wavelengths lags from the gamma flux for $82\pm32^\text{d}$~\citep{Rani14}, while the delay between the optical and gamma ranges is $\approx1.4^\text{d}$~\citep{Larionov13}.
These facts indicate that the optical emitting region is closer to the true jet base than the 15~GHz VLBI core and farther than the gamma-ray emitting region, which does not contradict the assumption about the synchrotron self-absorption action.
Additionally, \citet{Rani13} detected that the time lags between the S5 0716+714 variability at different radio frequencies follow power-law dependence, which extrapolation to optical range agrees with the observed time lag.

By analogy with VLBI jets, optical radiation can come from two regions with the optically thick and thin medium.
Unfortunately, the resolution of modern optical telescopes does not allow direct observation of these regions.
First, we assumed that most of the radiation comes from the optical core. 
In this case, the flux variability occurs due to Doppler factor changes in some of its parts.
We have obtained that $\delta_\text{var}$ changes from 7 to 14 without smooth transition from $\delta_\text{c}=5$.
This fact means that some directions of jet plasma motion do not realize. 
The presence of such selected directions has no natural physical explanation.
Second, the jet can contribute to the total radiation from the object. 
In this case, we assumed that the variability entirely forms in the compact optical core. 
If the electrons did not undergo significant radiation losses during propagation from the core to the jet ($\alpha_\text{pl}=\alpha_\text{c}$, Section~\ref{sec:4.2.1}), we obtained a change of $\delta_\text{var}$ from $\delta_\text{c}=5$ to 7.
The magnetic field is defined in the range from $10^2$ to $10^3$~G, except $\nu^\prime_1=9.5\cdot10^{13}$~Hz, for which it is impossible to estimate $B_\bot$ for filter pairs I---V, I---R.
Addition disadvantage of this case is the poor correspondence of the probability density function of the observed and theoretical fluxes. 
Note that these conclusions correspond to the assumption of both bright and faint jets. 
In the case in which jet electrons have significant radiation losses ($\alpha_\text{pl}=\alpha_\text{c}-0.5$, Section~\ref{sec:4.2.2}), the variable component Doppler factor is in the range of $\delta_\text{c}\leq\delta_\text{var}\leq8$ for bright and $5\leq\delta_\text{var}\leq11$ faint jets.  
The obtained magnetic field of values from 10 to $3\cdot10^2$~G agrees well with other independent estimates \citep{mag_field1, mag_field2}.
Therefore, this case is the most plausible.

The values of the Doppler factor used in our analysis are partly arbitrary.
For example, in the case of the single radiating region, we assume that the minimum flux of the constant component cannot be at large values of the Doppler factor. Therefore, we took the minimum $\delta_\text{c}=5$.
For higher $\delta_\text{c}$, the amplitude of the $\delta_\text{var}$ change holds, but the range of $\delta_\text{var}$ shifts to the higher values by $\delta_\text{c}$.
In the cases of two radiating regions, we used the average Doppler factor of the jet as the fixed value for $\delta_\text{c}$, which we assumed to be equal to 5 and 15 for more coverage of the values.
Regardless of the Doppler factor and the jet brightness in the source reference frame, the contributions from the jet and core are comparable in the observer's reference frame.
It confirms the presence of bright optical jets made from the analysis of the astrometric position shifts of the AGN nuclei between the observational data of VLBA and Gaia~\citep{Kovalev17, Kovalev20, Petrov17,Plavin19}.

The assumption of the S5~0716+714 jet has a region in which synchrotron self-absorption operates at optical frequencies, give a simple explanation for the observed long-term trend of color index changes under flux variability caused by geometric effects.
At the same time, the assumption of the physical process action in the jet has difficulties under an interpretation of the observed color index long-term trend.
First, it is necessary to ``adjust'' the parameters so that the color changes of the object are small in the bright state, whereas, with the flux decrease, the color changes would be more pronounced.
A possible solution to this problem is that, in the faint state, the object variability is caused only by physical processes. 
The object brightness is higher, the contribution to the variability from geometric effects is more significant.
Then, to produce the well-marked dependence in flux-flux diagrams, the physical processes leading to the brightness variability must cease to act at small angles of the radiating region velocity vector to the line of sight.
It's unbelievable.
Second, the parameters of the physical processes responsible for the flux variability must remain unchanged for at least 200 years in the reference frame comoving with relativistic plasma.
Therefore, our interpretation of the  blazar S5~0716+714 long-term variability is the most attractive.
Our basic assumption that part of the optical radiation forms in the optically thick jet medium can have some skepticism, because, as traditionally believed, the medium is transparent for optical radiation at a small distance from the true jet base.
Assuming the action of synchrotron self-absorption in a size region of one gravitational radius of a black hole with a mass of $(1-5)\cdot10^8M_\sun$, we estimated the magnetic field strength and the numerical density of the particles, which turned out to be acceptable values.
Under this, the radiating region size is $9\cdot10^{-5} - 4\cdot 10^{-4}$~pc, which is about 0.3$-$1.5~light days. 
This size seems small in the assumption that the observed radiation from gamma to optical frequencies is formed in the 43~GHz VLBI core representing a recollimation shock wave \citep[see, e.g. ][]{ DArcangelo07, Marscher08} because it locates far downstream from the true jet base.
However, this model does not explain the observed time delays in variability at different frequencies.
On the other hand, according to observations with a 2-minute time resolution, \citet{Raiteri21} found a minimum characteristic variability time scale of 0.2$^\text{d}$, which corresponds to the size in the source reference frame of less than 10$^{-3}$~pc.
\citet{Raiteri21} believe that this is the minimum size of the jet substructure.
But modeling of a high-energy hump in the blazar S5~0716+714 spectral energy distribution using simultaneous observational data during the 2015 flares resulted in the size of the emitting region of 0.02~pc \cite{MAGIC18}.
There is no contradiction with our analysis, in which the best agreement attains for the assumption of about two optical emitting regions, one of which is a compact region with synchrotron self-absorption, the other is an extended jet with an optically thin medium.
According to VLBI observations, the S5 0716+714 jet has a length of 5~mas ($\approx 23$~pc) at 15~GHz and about 100~mas ($\approx470$~pc) at 1.4~GHz \citep{PushkarevKovL17}.
But, the region with synchrotron self-absorption is very compact, even in the radio range.
According to the data of the \textit{RadioAstron} space VLBI at a frequency of 22~GHz, the upper limit on the size of the VLBI core is 12$\times$6~$\mu$as \citep{Kravchenko20a, Kravchenko20b}, which corresponds to $<$0.6~pc in the reference frame comoving relativistically moving plasma (for $\delta=10$).
For a conical geometry of the Blandford-K{\"o}nigl jet \citep{Blandford}, the approximation to optical frequencies gives the minimum size of the optical core $R_\text{opt}=R_{\text{22 GHz}}(\nu_\text{22 GHZ}/\nu_\text{opt})\approx 2\cdot10^{-5}$~pc (for the 22~GHz core angular size of 10~$\mu$as).
But if we assume that the transition of the jet shape from parabolic to conical \citep{Beskin17, KovPNokh20} occurs near the 22~GHz core, then $R_\text{opt}=3\cdot10^{-3}$~pc is proportional to the square root of the frequency ratio and is the upper limit.
We estimated the magnetic field using the value, which lies in the middle of this interval. 
\citet{Nokhrina20} determined that the width of the jet is 0.1$-$1~pc (for the mass of the central black hole of $<10^9$ solar masses) at the location of jet shape transition, i.e., at a distance from several tens of thousands to several hundred thousand parsecs from the true jet base.
The optical core locates near the jet base, its distance from the central black hole we can not estimate through our analysis. 
These facts make our assumption about the curved spectrum of the emitting region reasonable.
Additionally, in the framework of synchrotron self-absorption acting in the jet part, \citet{Butuzova21} explained the different color index behavior under intra-day optical variability and the fact that the larger frequency interval between two considered optical bands, the stronger BWB-chromatism is observed \citep{Dai15, Butuzova21}.

\section{Conclusion}

The properties of the blazar S5~0716+714 multi-band optical variability observed over several decades can be fully explained under the assumption of two radiating regions. One of the regions is a part of the jet in which the medium becomes transparent to optical radiation (optical core). A change in the Doppler factor of this region leads to the observed variability. The other emitting region is an extended, bright in the observer's reference frame jet with an optically thin medium. The electron energy spectra in the optical core and the jet are different: electrons emitting in the jet had undergone significant energy losses.

\section{Acknowlegments}

The authors are grateful to the anonymous Reviewers, whose recommendations allow us to make the analysis as complete as possible and the conclusions more convincing. This investigation was partly supported by the Russian Science Foundation grant 19-72-00105.

\bibliography{0716}{}

\begin{thebibliography}{}
\expandafter\ifx\csname natexlab\endcsname\relax\def\natexlab#1{#1}\fi
\providecommand{\url}[1]{\href{#1}{#1}}
\providecommand{\dodoi}[1]{doi:~\href{http://doi.org/#1}{\nolinkurl{#1}}}
\providecommand{\doeprint}[1]{\href{http://ascl.net/#1}{\nolinkurl{http://ascl.net/#1}}}
\providecommand{\doarXiv}[1]{\href{https://arxiv.org/abs/#1}{\nolinkurl{https://arxiv.org/abs/#1}}}

\bibitem[{{Agarwal} {et~al.}(2017){Agarwal}, {Mohan}, {Gupta}, {Mangalam},
  {Volvach}, {Aller}, {Aller}, {Gu}, {L{\"a}hteenm{\"a}ki}, {Tornikoski}, \&
  {Volvach}}]{Agarwal17}
{Agarwal}, A., {Mohan}, P., {Gupta}, A.~C., {et~al.} 2017, \mnras, 469, 813

\bibitem[{{Bach} {et~al.}(2005){Bach}, {Krichbaum}, {Ros}, {Britzen}, {Tian},
  {Kraus}, {Witzel}, \& {Zensus}}]{Bach05}
{Bach}, U., {Krichbaum}, T.~P., {Ros}, E., {et~al.} 2005, \aap, 433, 815

\bibitem[{{Beskin} {et~al.}(2017){Beskin}, {Chernoglazov}, {Kiselev}, \&
  {Nokhrina}}]{Beskin17}
{Beskin}, V.~S., {Chernoglazov}, A.~V., {Kiselev}, A.~M., \& {Nokhrina}, E.~E.
  2017, \mnras, 472, 3971

\bibitem[{{Bhatta} {et~al.}(2013){Bhatta}, {Webb}, {Hollingsworth}, {Dhalla},
  {Khanuja}, {Bachev}, {Blinov}, {B{\"o}ttcher}, {Bravo Calle}, {Calcidese},
  {Capezzali}, {Carosati}, {Chigladze}, {Collins}, {Coloma}, {Efimov}, {Gupta},
  {Hu}, {Kurtanidze}, {Lamerato}, {Larionov}, {Lee}, {Lindfors}, {Murphy},
  {Nilsson}, {Ohlert}, {Oksanen}, {P{\"a}{\"a}kk{\"o}nen}, {Pollock}, {Rani},
  {Reinthal}, {Rodriguez}, {Ros}, {Roustazadeh}, {Sagar}, {Sanchez}, {Shastri},
  {Sillanp{\"a}{\"a}}, {Strigachev}, {Takalo}, {Vennes}, {Villata},
  {Villforth}, {Wu}, \& {Zhou}}]{Bhatta13}
{Bhatta}, G., {Webb}, J.~R., {Hollingsworth}, H., {et~al.} 2013, \aap, 558, A92

\bibitem[{{Blandford} \& {K{\"o}nigl}(1979)}]{Blandford}
{Blandford}, R.~D., \& {K{\"o}nigl}, A. 1979, \apj, 232, 34

\bibitem[{{Britzen} {et~al.}(2009){Britzen}, {Kam}, {Witzel}, {Agudo}, {Aller},
  {Aller}, {Karouzos}, {Eckart}, \& {Zensus}}]{Britzen09}
{Britzen}, S., {Kam}, V.~A., {Witzel}, A., {et~al.} 2009, \aap, 508, 1205

\bibitem[{{Butuzova}(2018{\natexlab{a}})}]{Butuzova18a}
{Butuzova}, M.~S. 2018{\natexlab{a}}, Astronomy Reports, 62, 116

\bibitem[{{Butuzova}(2018{\natexlab{b}})}]{Butuzova18b}
---. 2018{\natexlab{b}}, Astronomy Reports, 62, 654

\bibitem[{{Butuzova}(2021)}]{Butuzova21}
---. 2021, Astroparticle Physics, 129, 102577

\bibitem[{{Butuzova} \& {Pushkarev}(2020)}]{Butuzova2020jet}
{Butuzova}, M.~S., \& {Pushkarev}, A.~B. 2020, Universe, 6, 191

\bibitem[{{Bychkova} {et~al.}(2015){Bychkova}, {Vol'vach}, {Kardashev},
  {Larionov}, {Vlasyuk}, {Spiridonova}, {Vol'vach}, {L{\"a}hteenm{\"a}ki},
  {Tornikoski}, {Aller}, \& {Aller}}]{Bychkova15}
{Bychkova}, V.~S., {Vol'vach}, A.~E., {Kardashev}, N.~S., {et~al.} 2015,
  Astronomy Reports, 59, 851

\bibitem[{{Camenzind} \& {Krockenberger}(1992)}]{CamKrock92}
{Camenzind}, M., \& {Krockenberger}, M. 1992, \aap, 255, 59

\bibitem[{{Choloniewski}(1981)}]{Choloniewski81}
{Choloniewski}, J. 1981, \actaa, 31, 293

\bibitem[{{Dai} {et~al.}(2015){Dai}, {Zeng}, {Jiang}, {Fan}, {Hu}, {Zhang},
  {Yang}, {Yan}, {Wang}, \& {Zhang}}]{Dai15}
{Dai}, B.-z., {Zeng}, W., {Jiang}, Z.-j., {et~al.} 2015, \apjs, 218, 18

\bibitem[{{D'Arcangelo} {et~al.}(2007){D'Arcangelo}, {Marscher}, {Jorstad},
  {Smith}, {Larionov}, {Hagen-Thorn}, {Kopatskaya}, {Williams}, \&
  {Gear}}]{DArcangelo07}
{D'Arcangelo}, F.~D., {Marscher}, A.~P., {Jorstad}, S.~G., {et~al.} 2007,
  \apjl, 659, L107

\bibitem[{{Donnarumma} {et~al.}(2009){Donnarumma}, {Vittorini}, {Vercellone},
  {del Monte}, {Feroci}, {D'Ammando}, {Pacciani}, {Chen}, {Tavani},
  {Bulgarelli}, {Giuliani}, {Longo}, {Pucella}, {Argan}, {Barbiellini},
  {Boffelli}, {Caraveo}, {Cattaneo}, {Cocco}, {Costa}, {DeParis}, {Di Cocco},
  {Evangelista}, {Fiorini}, {Froysland}, {Frutti}, {Fuschino}, {Galli},
  {Gianotti}, {Labanti}, {Lapshov}, {Lazzarotto}, {Lipari}, {Marisaldi},
  {Mastropietro}, {Mereghetti}, {Morelli}, {Morselli}, {Pellizzoni}, {Perotti},
  {Picozza}, {Porrovecchio}, {Prest}, {Rapisarda}, {Rappoldi}, {Rubini},
  {Soffitta}, {Trifoglio}, {Trois}, {Vallazza}, {Zambra}, {Zanello}, {Pittori},
  {Santolamazza}, {Verrecchia}, {Giommi}, {Colafrancesco}, {Salotti},
  {Villata}, {Raiteri}, {Chen}, {Efimova}, {Jordan}, {Konstantinova},
  {Koptelova}, {Kurtanidze}, {Larionov}, {Ros}, {Sadun}, {Anderhub},
  {Antonelli}, {Antoranz}, {Backes}, {Baixeras}, {Balestra}, {Barrio},
  {Bartko}, {Bastieri}, {Gonz{\'a}lez}, {Becker}, {Bednarek}, {Berger},
  {Bernardini}, {Biland}, {Bock}, {Bonnoli}, {Bordas}, {Tridon}, {Bosch-Ramon},
  {Bretz}, {Britvitch}, {Camara}, {Carmona}, {Chilingarian}, {Commichau},
  {Contreras}, {Cortina}, {Costado}, {Covino}, {Curtef}, {Dazzi}, {DeAngelis},
  {DeCea del Pozo}, {de los Reyes}, {DeLotto}, {DeMaria}, {DeSabata}, {Mendez},
  {Dominguez}, {Dorner}, {Doro}, {Elsaesser}, {Errando}, {Ferenc},
  {Fern{\'a}ndez}, {Firpo}, {Fonseca}, {Font}, {Galante}, {Garc{\'\i}a
  L{\'o}pez}, {Garczarczyk}, {Gaug}, {Goebel}, {Hadasch}, {Hayashida},
  {Herrero}, {H{\"o}hne-M{\"o}nch}, {Hose}, {Hsu}, {Huber}, {Jogler},
  {Kranich}, {La Barbera}, {Laille}, {Leonardo}, {Lindfors}, {Lombardi},
  {L{\'o}pez}, {Lorenz}, {Majumdar}, {Maneva}, {Mankuzhiyil}, {Mannheim},
  {Maraschi}, {Mariotti}, {Mart{\'\i}nez}, {Mazin}, {Meucci}, {Meyer},
  {Miranda}, {Mirzoyan}, {Mold{\'o}n}, {Moles}, {Moralejo}, {Nieto}, {Nilsson},
  {Ninkovic}, {Oya}, {Paoletti}, {Paredes}, {Pasanen}, {Pascoli}, {Pauss},
  {Pegna}, {Perez-Torres}, {Persic}, {Peruzzo}, {Prada}, {Prandini},
  {Puchades}, {Raymers}, {Rhode}, {Rib{\'o}}, {Rico}, {Rissi}, {Robert},
  {R{\"u}gamer}, {Saggion}, {Saito}, {Salvati}, {Sanchez-Conde}, {Sartori},
  {Satalecka}, {Scalzotto}, {Scapin}, {Schweizer}, {Shayduk}, {Shinozaki},
  {Shore}, {Sidro}, {Sierpowska-Bartosik}, {Sillanp{\"a}{\"a}}, {Sitarek},
  {Sobczynska}, {Spanier}, {Stamerra}, {Stark}, {Takalo}, {Tavecchio},
  {Temnikov}, {Tescaro}, {Teshima}, {Tluczykont}, {Torres}, {Turini}, {Vankov},
  {Venturini}, {Vitale}, {Wagner}, {Wittek}, {Zabalza}, {Zandanel}, {Zanin},
  {Zapatero}, {Acciari}, {Aliu}, {Arlen}, {Beilicke}, {Benbow}, {Bradbury},
  {Buckley}, {Bugaev}, {Butt}, {Byrum}, {Cannon}, {Cesarini}, {Chow}, {Ciupik},
  {Cogan}, {Colin}, {Cui}, {Daniel}, {Dickherber}, {Duke}, {Ergin}, {Fegan},
  {Finley}, {Finnegan}, {Fortin}, {Furniss}, {Gall}, {Gillanders}, {Guenette},
  {Gyuk}, {Grube}, {Hanna}, {Holder}, {Horan}, {Hui}, {Humensky}, {Imran},
  {Kaaret}, {Karlsson}, {Kertzman}, {Kieda}, {Kildea}, {Konopelko},
  {Krawczynski}, {Krennrich}, {Lang}, {LeBohec}, {Maier}, {McCann},
  {McCutcheon}, {Milovanovic}, {Moriarty}, {Nagai}, {Ong}, {Otte}, {Pandel},
  {Perkins}, {Pichel}, {Pohl}, {Ragan}, {Reyes}, {Reynolds}, {Roache}, {Rose},
  {Schroedter}, {Sembroski}, {Smith}, {Steele}, {Swordy}, {Theiling}, {Toner},
  {Valcarcel}, {Varlotta}, {Wakely}, {Ward}, {Weekes}, {Weinstein}, {Williams},
  {Wissel}, {Wood}, \& {Zitzer}}]{Donnarumma09}
{Donnarumma}, I., {Vittorini}, V., {Vercellone}, S., {et~al.} 2009, \apjl, 691,
  L13

\bibitem[{{Doroshenko} {et~al.}(2005){Doroshenko}, {Sergeev}, {Merkulova},
  {Sergeeva}, {Golubinsky}, {Pronik}, \& {Okhmat}}]{Doroshenko2005}
{Doroshenko}, V.~T., {Sergeev}, S.~G., {Merkulova}, N.~I., {et~al.} 2005,
  Astrophysics, 48, 156

\bibitem[{{Feng} {et~al.}(2020){Feng}, {Yang}, {Yang}, {Liu}, {Bai}, {Li},
  {Zhao}, {Zhang}, {Li}, {Xiao}, {Xin}, {Xing}, {Lu}, {Xu}, {Wang}, {Wang},
  {Zhang}, {Zhang}, {Lun}, \& {He}}]{Feng20}
{Feng}, H.-C., {Yang}, S., {Yang}, Z.-X., {et~al.} 2020, \apj, 902, 42

\bibitem[{{Field} \& {Rogers}(1993)}]{mag_field1}
{Field}, G.~B., \& {Rogers}, R.~D. 1993, \apj, 403, 94

\bibitem[{{Ghisellini} {et~al.}(1997){Ghisellini}, {Villata}, {Raiteri},
  {Bosio}, {de Francesco}, {Latini}, {Maesano}, {Massaro}, {Montagni}, {Nesci},
  {Tosti}, {Fiorucci}, {Pian}, {Maraschi}, {Treves}, {Comastri}, \&
  {Mignoli}}]{Ghisellini97}
{Ghisellini}, G., {Villata}, M., {Raiteri}, C.~M., {et~al.} 1997, \aap, 327, 61

\bibitem[{{Gu} {et~al.}(2006){Gu}, {Lee}, {Pak}, {Yim}, \&
  {Fletcher}}]{GuLee06}
{Gu}, M.~F., {Lee}, C.~U., {Pak}, S., {Yim}, H.~S., \& {Fletcher}, A.~B. 2006,
  \aap, 450, 39

\bibitem[{{Hagen-Thorn}(1997)}]{Hagen-Thorn97}
{Hagen-Thorn}, V.~A. 1997, Astronomy Letters, 23, 19

\bibitem[{{Homan} {et~al.}(2001){Homan}, {Ojha}, {Wardle}, {Roberts}, {Aller},
  {Aller}, \& {Hughes}}]{Homan01}
{Homan}, D.~C., {Ojha}, R., {Wardle}, J. F.~C., {et~al.} 2001, \apj, 549, 840

\bibitem[{{Hong} {et~al.}(2017){Hong}, {Xiong}, \& {Bai}}]{Hong17}
{Hong}, S., {Xiong}, D., \& {Bai}, J. 2017, \aj, 154, 42

\bibitem[{{Isler} {et~al.}(2017){Isler}, {Urry}, {Coppi}, {Bailyn}, {Brady},
  {MacPherson}, {Buxton}, \& {Hasan}}]{Isler17}
{Isler}, J.~C., {Urry}, C.~M., {Coppi}, P., {et~al.} 2017, \apj, 844, 107

\bibitem[{{Kaur} {et~al.}(2018){Kaur}, {Baliyan}, {Chandra}, {Sameer}, \&
  {Ganesh}}]{Kaur18}
{Kaur}, N., {Baliyan}, K.~S., {Chandra}, S., {Sameer}, \& {Ganesh}, S. 2018,
  \aj, 156, 36

\bibitem[{{Kim} {et~al.}(2020){Kim}, {Krichbaum}, {Broderick}, {Wielgus},
  {Blackburn}, {G{\'o}mez}, {Johnson}, \& {Bouman}}]{Kim2020}
{Kim}, J.-Y., {Krichbaum}, T.~P., {Broderick}, A.~E., {et~al.} 2020, \aap, 640,
  A69

\bibitem[{{Kovalev} {et~al.}(2017){Kovalev}, {Petrov}, \& {Plavin}}]{Kovalev17}
{Kovalev}, Y.~Y., {Petrov}, L., \& {Plavin}, A.~V. 2017, \aap, 598, L1

\bibitem[{{Kovalev} {et~al.}(2020{\natexlab{a}}){Kovalev}, {Pushkarev},
  {Nokhrina}, {Plavin}, {Beskin}, {Chernoglazov}, {Lister}, \&
  {Savolainen}}]{KovPNokh20}
{Kovalev}, Y.~Y., {Pushkarev}, A.~B., {Nokhrina}, E.~E., {et~al.}
  2020{\natexlab{a}}, \mnras, 495, 3576

\bibitem[{{Kovalev} {et~al.}(2020{\natexlab{b}}){Kovalev}, {Zobnina}, {Plavin},
  \& {Blinov}}]{Kovalev20}
{Kovalev}, Y.~Y., {Zobnina}, D.~I., {Plavin}, A.~V., \& {Blinov}, D.
  2020{\natexlab{b}}, \mnras, 493, L54

\bibitem[{{Kravchenko} {et~al.}(2020{\natexlab{a}}){Kravchenko}, {G{\'o}mez},
  {Kovalev}, \& {Voitsik}}]{Kravchenko20a}
{Kravchenko}, E.~V., {G{\'o}mez}, J.~L., {Kovalev}, Y.~Y., \& {Voitsik}, P.~A.
  2020{\natexlab{a}}, Advances in Space Research, 65, 720

\bibitem[{{Kravchenko} {et~al.}(2020{\natexlab{b}}){Kravchenko}, {G{\'o}mez},
  {Kovalev}, {Lobanov}, {Savolainen}, {Bruni}, {Fuentes}, {Anderson},
  {Jorstad}, {Marscher}, {Tornikoski}, {L{\"a}hteenm{\"a}ki}, \&
  {Lisakov}}]{Kravchenko20b}
{Kravchenko}, E.~V., {G{\'o}mez}, J.~L., {Kovalev}, Y.~Y., {et~al.}
  2020{\natexlab{b}}, \apj, 893, 68

\bibitem[{{Larionov} {et~al.}(2008){Larionov}, {Jorstad}, {Marscher},
  {Raiteri}, {Villata}, {Agudo}, {Aller}, {Arkharov}, {Asfandiyarov}, {Bach},
  {Bachev}, {Berdyugin}, {B{\"o}ttcher}, {Buemi}, {Calcidese}, {Carosati},
  {Charlot}, {Chen}, {di Paola}, {Dolci}, {Dogru}, {Doroshenko}, {Efimov},
  {Erdem}, {Frasca}, {Fuhrmann}, {Giommi}, {Glowienka}, {Gupta}, {Gurwell},
  {Hagen-Thorn}, {Hsiao}, {Ibrahimov}, {Jordan}, {Kamada}, {Konstantinova},
  {Kopatskaya}, {Kovalev}, {Kovalev}, {Kurtanidze}, {L{\"a}hteenm{\"a}ki},
  {Lanteri}, {Larionova}, {Leto}, {Le Campion}, {Lee}, {Lindfors}, {Marilli},
  {McHardy}, {Mingaliev}, {Nazarov}, {Nieppola}, {Nilsson}, {Ohlert},
  {Pasanen}, {Porter}, {Pursimo}, {Ros}, {Sadakane}, {Sadun}, {Sergeev},
  {Smith}, {Strigachev}, {Sumitomo}, {Takalo}, {Tanaka}, {Trigilio}, {Umana},
  {Ungerechts}, {Volvach}, \& {Yuan}}]{Larionov2008}
{Larionov}, V.~M., {Jorstad}, S.~G., {Marscher}, A.~P., {et~al.} 2008, \aap,
  492, 389

\bibitem[{{Larionov} {et~al.}(2013){Larionov}, {Jorstad}, {Marscher},
  {Morozova}, {Blinov}, {Hagen-Thorn}, {Konstantinova}, {Kopatskaya},
  {Larionova}, {Larionova}, \& {Troitsky}}]{Larionov13}
---. 2013, \apj, 768, 40

\bibitem[{{Liao} {et~al.}(2014){Liao}, {Bai}, {Liu}, {Weng}, {Chen}, \&
  {Li}}]{Liao14}
{Liao}, N.~H., {Bai}, J.~M., {Liu}, H.~T., {et~al.} 2014, \apj, 783, 83

\bibitem[{{Liodakis} \& {Pavlidou}(2015)}]{Liodakis15}
{Liodakis}, I., \& {Pavlidou}, V. 2015, \mnras, 451, 2434

\bibitem[{{Lister} {et~al.}(2013){Lister}, {Aller}, {Aller}, {Homan},
  {Kellermann}, {Kovalev}, {Pushkarev}, {Richards}, {Ros}, \&
  {Savolainen}}]{Lister13}
{Lister}, M.~L., {Aller}, M.~F., {Aller}, H.~D., {et~al.} 2013, \aj, 146, 120

\bibitem[{{Lister} {et~al.}(2016){Lister}, {Aller}, {Aller}, {Homan},
  {Kellermann}, {Kovalev}, {Pushkarev}, {Richards}, {Ros}, \&
  {Savolainen}}]{Lister16}
---. 2016, \aj, 152, 12

\bibitem[{{Liu} {et~al.}(2019){Liu}, {Feng}, {Xin}, {Bai}, {Li}, \&
  {Wang}}]{Liu2019}
{Liu}, H.~T., {Feng}, H.~C., {Xin}, Y.~X., {et~al.} 2019, \apj, 880, 155

\bibitem[{{Lobanov}(1998)}]{Lobanov98}
{Lobanov}, A.~P. 1998, \aap, 330, 79

\bibitem[{{Lyutikov} {et~al.}(2005){Lyutikov}, {Pariev}, \&
  {Gabuzda}}]{Lyutikov05}
{Lyutikov}, M., {Pariev}, V.~I., \& {Gabuzda}, D.~C. 2005, \mnras, 360, 869

\bibitem[{{MAGIC Collaboration} {et~al.}(2018){MAGIC Collaboration}, {Ahnen},
  {Ansoldi}, {Antonelli}, {Arcaro}, {Baack}, {Babi{\'c}}, {Banerjee},
  {Bangale}, {Barres de Almeida}, {Barrio}, {Becerra Gonz{\'a}lez}, {Bednarek},
  {Bernardini}, {Ch Berse}, {Berti}, {Bhattacharyya}, {Biland}, {Blanch},
  {Bonnoli}, {Carosi}, {Carosi}, {Ceribella}, {Chatterjee}, {Colak}, {Colin},
  {Colombo}, {Contreras}, {Cortina}, {Covino}, {Cumani}, {da Vela}, {Dazzi},
  {de Angelis}, {de Lotto}, {Delfino}, {Delgado}, {di Pierro},
  {Dom{\'\i}nguez}, {Dominis Prester}, {Dorner}, {Doro}, {Einecke},
  {Elsaesser}, {Fallah Ramazani}, {Fern{\'a}ndez-Barral}, {Fidalgo}, {Fonseca},
  {Font}, {Fruck}, {Galindo}, {Gallozzi}, {Garc{\'\i}a L{\'o}pez},
  {Garczarczyk}, {Gaug}, {Giammaria}, {Godinovi{\'c}}, {Gora}, {Guberman},
  {Hadasch}, {Hahn}, {Hassan}, {Hayashida}, {Herrera}, {Hose}, {Hrupec},
  {Ishio}, {Konno}, {Kubo}, {Kushida}, {Kuve{\v{z}}di{\'c}}, {Lelas},
  {Lindfors}, {Lombardi}, {Longo}, {L{\'o}pez}, {Maggio}, {Majumdar},
  {Makariev}, {Maneva}, {Manganaro}, {Mannheim}, {Maraschi}, {Mariotti},
  {Mart{\'\i}nez}, {Masuda}, {Mazin}, {Mielke}, {Minev}, {Miranda}, {Mirzoyan},
  {Moralejo}, {Moreno}, {Moretti}, {Nagayoshi}, {Neustroev}, {Niedzwiecki},
  {Nievas Rosillo}, {Nigro}, {Nilsson}, {Ninci}, {Nishijima}, {Noda},
  {Nogu{\'e}s}, {Paiano}, {Palacio}, {Paneque}, {Paoletti}, {Paredes},
  {Pedaletti}, {Peresano}, {Persic}, {Prada Moroni}, {Prandini}, {Puljak},
  {Garcia}, {Reichardt}, {Rhode}, {Rib{\'o}}, {Rico}, {Righi}, {Rugliancich},
  {Saito}, {Satalecka}, {Schweizer}, {Sitarek}, {{\v{S}}nidari{\'c}},
  {Sobczynska}, {Stamerra}, {Strzys}, {Suri{\'c}}, {Takahashi}, {Takalo},
  {Tavecchio}, {Temnikov}, {Terzi{\'c}}, {Teshima}, {Torres-Alb{\`a}},
  {Treves}, {Tsujimoto}, {Vanzo}, {Vazquez Acosta}, {Vovk}, {Ward}, {Will},
  {Zari{\'c}}, {Fermi-Lat Collaboration}, {Bastieri}, {Gasparrini}, {Lott},
  {Rani}, {Thompson}, {MWL Collaborators}, {Agudo}, {Angelakis}, {Borman},
  {Casadio}, {Grishina}, {Gurwell}, {Hovatta}, {Itoh}, {J{\"a}rvel{\"a}},
  {Jermak}, {Jorstad}, {Kopatskaya}, {Kraus}, {Krichbaum}, {Kuin},
  {L{\"a}hteenm{\"a}ki}, {Larionov}, {Larionova}, {Lien}, {Madejski},
  {Marscher}, {Myserlis}, {Max-Moerbeck}, {Molina}, {Morozova}, {Nalewajko},
  {Pearson}, {Ramakrishnan}, {Readhead}, {Reeves}, {Savchenko}, {Steele},
  {Tornikoski}, {Troitskaya}, {Troitsky}, {Vasilyev}, \& {Zensus}}]{MAGIC18}
{MAGIC Collaboration}, {Ahnen}, M.~L., {Ansoldi}, S., {et~al.} 2018, \aap, 619,
  A45

\bibitem[{{Marscher} \& {Gear}(1985)}]{MarscherGear85}
{Marscher}, A.~P., \& {Gear}, W.~K. 1985, \apj, 298, 114

\bibitem[{{Marscher} {et~al.}(2008){Marscher}, {Jorstad}, {D'Arcangelo},
  {Smith}, {Williams}, {Larionov}, {Oh}, {Olmstead}, {Aller}, {Aller},
  {McHardy}, {L{\"a}hteenm{\"a}ki}, {Tornikoski}, {Valtaoja}, {Hagen-Thorn},
  {Kopatskaya}, {Gear}, {Tosti}, {Kurtanidze}, {Nikolashvili}, {Sigua},
  {Miller}, \& {Ryle}}]{Marscher08}
{Marscher}, A.~P., {Jorstad}, S.~G., {D'Arcangelo}, F.~D., {et~al.} 2008, \nat,
  452, 966

\bibitem[{{Massaro} {et~al.}(2004{\natexlab{a}}){Massaro}, {Perri}, {Giommi},
  \& {Nesci}}]{Massaro04a}
{Massaro}, E., {Perri}, M., {Giommi}, P., \& {Nesci}, R. 2004{\natexlab{a}},
  \aap, 413, 489

\bibitem[{{Massaro} {et~al.}(2004{\natexlab{b}}){Massaro}, {Perri}, {Giommi},
  {Nesci}, \& {Verrecchia}}]{Massaro04b}
{Massaro}, E., {Perri}, M., {Giommi}, P., {Nesci}, R., \& {Verrecchia}, F.
  2004{\natexlab{b}}, \aap, 422, 103

\bibitem[{{Mead} {et~al.}(1990){Mead}, {Ballard}, {Brand}, {Hough}, {Brindle},
  \& {Bailey}}]{Mead90}
{Mead}, A.~R.~G., {Ballard}, K.~R., {Brand}, P.~W.~J.~L., {et~al.} 1990, \aaps,
  83, 183

\bibitem[{{Mondal} \& {Mukhopadhyay}(2019)}]{mag_field2}
{Mondal}, T., \& {Mukhopadhyay}, B. 2019, \mnras, 486, 3465

\bibitem[{{Nieppola} {et~al.}(2006){Nieppola}, {Tornikoski}, \&
  {Valtaoja}}]{Nieppola06}
{Nieppola}, E., {Tornikoski}, M., \& {Valtaoja}, E. 2006, \aap, 445, 441

\bibitem[{{Nilsson} {et~al.}(2008){Nilsson}, {Pursimo}, {Sillanp{\"a}{\"a}},
  {Takalo}, \& {Lindfors}}]{Nilsson08}
{Nilsson}, K., {Pursimo}, T., {Sillanp{\"a}{\"a}}, A., {Takalo}, L.~O., \&
  {Lindfors}, E. 2008, \aap, 487, L29

\bibitem[{{Nokhrina} {et~al.}(2020){Nokhrina}, {Kovalev}, \&
  {Pushkarev}}]{Nokhrina20}
{Nokhrina}, E.~E., {Kovalev}, Y.~Y., \& {Pushkarev}, A.~B. 2020, \mnras, 498,
  2532

\bibitem[{{Pacholczyk}(1970)}]{Pachol}
{Pacholczyk}, A.~G. 1970, {Radio astrophysics. Nonthermal processes in galactic
  and extragalactic sources} (San Francisco: Freeman)

\bibitem[{{Petrov} \& {Kovalev}(2017)}]{Petrov17}
{Petrov}, L., \& {Kovalev}, Y.~Y. 2017, \mnras, 467, L71

\bibitem[{{Pian} {et~al.}(2005){Pian}, {Falomo}, \& {Treves}}]{Pian05}
{Pian}, E., {Falomo}, R., \& {Treves}, A. 2005, \mnras, 361, 919

\bibitem[{{Plavin} {et~al.}(2019){Plavin}, {Kovalev}, \& {Petrov}}]{Plavin19}
{Plavin}, A.~V., {Kovalev}, Y.~Y., \& {Petrov}, L.~Y. 2019, \apj, 871, 143

\bibitem[{{Poon} {et~al.}(2009){Poon}, {Fan}, \& {Fu}}]{Poon09}
{Poon}, H., {Fan}, J.~H., \& {Fu}, J.~N. 2009, \apjs, 185, 511

\bibitem[{{Pursimo} {et~al.}(2002){Pursimo}, {Nilsson}, {Takalo},
  {Sillanp{\"a}{\"a}}, {Heidt}, \& {Pietil{\"a}}}]{Pursimo02}
{Pursimo}, T., {Nilsson}, K., {Takalo}, L.~O., {et~al.} 2002, \aap, 381, 810

\bibitem[{{Pushkarev} {et~al.}(2012){Pushkarev}, {Hovatta}, {Kovalev},
  {Lister}, {Lobanov}, {Savolainen}, \& {Zensus}}]{Pushkarev2012}
{Pushkarev}, A.~B., {Hovatta}, T., {Kovalev}, Y.~Y., {et~al.} 2012, \aap, 545,
  A113

\bibitem[{{Pushkarev} {et~al.}(2017){Pushkarev}, {Kovalev}, {Lister}, \&
  {Savolainen}}]{PushkarevKovL17}
{Pushkarev}, A.~B., {Kovalev}, Y.~Y., {Lister}, M.~L., \& {Savolainen}, T.
  2017, \mnras, 468, 4992

\bibitem[{{Raiteri} {et~al.}(2003){Raiteri}, {Villata}, {Tosti}, {Nesci},
  {Massaro}, {Aller}, {Aller}, {Ter{\"a}sranta}, {Kurtanidze}, {Nikolashvili},
  {Ibrahimov}, {Papadakis}, {Krichbaum}, {Kraus}, {Witzel}, {Ungerechts},
  {Lisenfeld}, {Bach}, {Cim{\`o}}, {Ciprini}, {Fuhrmann}, {Kimeridze},
  {Lanteri}, {Maesano}, {Montagni}, {Nucciarelli}, \& {Ostorero}}]{Raiteri03}
{Raiteri}, C.~M., {Villata}, M., {Tosti}, G., {et~al.} 2003, \aap, 402, 151

\bibitem[{{Raiteri} {et~al.}(2021){Raiteri}, {Villata}, {Carosati},
  {Ben{\'\i}tez}, {Kurtanidze}, {Gupta}, {Mirzaqulov}, {D'Ammando}, {Larionov},
  {Pursimo}, {Acosta-Pulido}, {Baida}, {Balmaverde}, {Bonnoli}, {Borman},
  {Carnerero}, {Chen}, {Dhiman}, {Di Maggio}, {Ehgamberdiev}, {Hiriart},
  {Kimeridze}, {Kurtanidze}, {Lin}, {Lopez}, {Marchini}, {Matsumoto}, {Mujica},
  {Nakamura}, {Nikiforova}, {Nikolashvili}, {Okhmat}, {Otero-Santos}, {Rizzi},
  {Sakamoto}, {Semkov}, {Sigua}, {Stiaccini}, {Troitsky}, {Tsai}, {Vasilyev},
  \& {Zhovtan}}]{Raiteri21}
{Raiteri}, C.~M., {Villata}, M., {Carosati}, D., {et~al.} 2021, \mnras, 501,
  1100

\bibitem[{{Rani} {et~al.}(2015){Rani}, {Krichbaum}, {Marscher}, {Hodgson},
  {Fuhrmann}, {Angelakis}, {Britzen}, \& {Zensus}}]{Rani15}
{Rani}, B., {Krichbaum}, T.~P., {Marscher}, A.~P., {et~al.} 2015, \aap, 578,
  A123

\bibitem[{{Rani} {et~al.}(2014){Rani}, {Krichbaum}, {Marscher}, {Jorstad},
  {Hodgson}, {Fuhrmann}, \& {Zensus}}]{Rani14}
---. 2014, \aap, 571, L2

\bibitem[{{Rani} {et~al.}(2013){Rani}, {Krichbaum}, {Fuhrmann}, {B{\"o}ttcher},
  {Lott}, {Aller}, {Aller}, {Angelakis}, {Bach}, {Bastieri}, {Falcone},
  {Fukazawa}, {Gabanyi}, {Gupta}, {Gurwell}, {Itoh}, {Kawabata}, {Krips},
  {L{\"a}hteenm{\"a}ki}, {Liu}, {Marchili}, {Max-Moerbeck}, {Nestoras},
  {Nieppola}, {Quintana-Lacaci}, {Readhead}, {Richards}, {Sasada}, {Sievers},
  {Sokolovsky}, {Stroh}, {Tammi}, {Tornikoski}, {Uemura}, {Ungerechts},
  {Urano}, \& {Zensus}}]{Rani13}
{Rani}, B., {Krichbaum}, T.~P., {Fuhrmann}, L., {et~al.} 2013, \aap, 552, A11

\bibitem[{{Rastorgueva} {et~al.}(2009){Rastorgueva}, {Wiik}, {Savolainen},
  {Takalo}, {Valtaoja}, {Vetukhnovskaya}, \& {Sokolovsky}}]{Rastorgueva09}
{Rastorgueva}, E.~A., {Wiik}, K., {Savolainen}, T., {et~al.} 2009, \aap, 494,
  L5

\bibitem[{{Rastorgueva} {et~al.}(2011{\natexlab{a}}){Rastorgueva}, {Wiik},
  {Bajkova}, {Valtaoja}, {Takalo}, {Vetukhnovskaya}, \&
  {Mahmud}}]{Rastorgueva11}
{Rastorgueva}, E.~A., {Wiik}, K.~J., {Bajkova}, A.~T., {et~al.}
  2011{\natexlab{a}}, \aap, 529, A2

\bibitem[{{Rastorgueva} {et~al.}(2011{\natexlab{b}}){Rastorgueva}, {Wiik},
  {Bajkova}, {Valtaoja}, {Takalo}, {Vetukhnovskaya}, \&
  {Mahmud}}]{Rastorgueva2011}
---. 2011{\natexlab{b}}, \aap, 529, A2

\bibitem[{{Sandage}(1973)}]{Sandage73}
{Sandage}, A. 1973, \apj, 183, 711

\bibitem[{{Sbarrato} {et~al.}(2012){Sbarrato}, {Ghisellini}, {Maraschi}, \&
  {Colpi}}]{Sbarrato12}
{Sbarrato}, T., {Ghisellini}, G., {Maraschi}, L., \& {Colpi}, M. 2012, \mnras,
  421, 1764

\bibitem[{{Sbarufatti} {et~al.}(2005){Sbarufatti}, {Treves}, \&
  {Falomo}}]{Sbarufatti05}
{Sbarufatti}, B., {Treves}, A., \& {Falomo}, R. 2005, \apj, 635, 173

\bibitem[{{Schlafly} \& {Finkbeiner}(2011)}]{Ext}
{Schlafly}, E.~F., \& {Finkbeiner}, D.~P. 2011, \apj, 737, 103

\bibitem[{{Sergeev} {et~al.}(2005){Sergeev}, {Doroshenko}, {Golubinskiy},
  {Merkulova}, \& {Sergeeva}}]{Sergeev2005}
{Sergeev}, S.~G., {Doroshenko}, V.~T., {Golubinskiy}, Y.~V., {Merkulova},
  N.~I., \& {Sergeeva}, E.~A. 2005, \apj, 622, 129

\bibitem[{{Slish}(1963)}]{Slish}
{Slish}, V.~I. 1963, \nat, 199, 682

\bibitem[{{Stalin} {et~al.}(2009){Stalin}, {Kawabata}, {Uemura}, {Yoshida},
  {Kawai}, {Yanagisawa}, {Shimizu}, {Kuroda}, {Nagayama}, \& {Toda}}]{Stalin09}
{Stalin}, C.~S., {Kawabata}, K.~S., {Uemura}, M., {et~al.} 2009, \mnras, 399,
  1357

\bibitem[{{Tanihata} {et~al.}(2004){Tanihata}, {Kataoka}, {Takahashi}, \&
  {Madejski}}]{Tanihata04}
{Tanihata}, C., {Kataoka}, J., {Takahashi}, T., \& {Madejski}, G.~M. 2004,
  \apj, 601, 759

\bibitem[{{Tramacere} {et~al.}(2007){Tramacere}, {Giommi}, {Massaro}, {Perri},
  {Nesci}, {Colafrancesco}, {Tagliaferri}, {Chincarini}, {Falcone}, {Burrows},
  {Roming}, {McMath Chester}, \& {Gehrels}}]{Tramacere07}
{Tramacere}, A., {Giommi}, P., {Massaro}, E., {et~al.} 2007, \aap, 467, 501

\bibitem[{{Villata} {et~al.}(1998){Villata}, {Raiteri}, {Lanteri}, {Sobrito},
  \& {Cavallone}}]{Villata1998}
{Villata}, M., {Raiteri}, C.~M., {Lanteri}, L., {Sobrito}, G., \& {Cavallone},
  M. 1998, \aaps, 130, 305

\bibitem[{{Volvach} {et~al.}(2012){Volvach}, {Volvach}, {Bychkova},
  {Kardashev}, {Larionov}, {Vlasjuk}, {Spiridonova}, {Lachteenmaki},
  {Tornikoski}, {Nieppola}, {Aller}, \& {Aller}}]{Volvach12}
{Volvach}, A.~E., {Volvach}, L.~N., {Bychkova}, V.~S., {et~al.} 2012, Astronomy
  Reports, 56, 275

\bibitem[{{Wagner} {et~al.}(1996){Wagner}, {Witzel}, {Heidt}, {Krichbaum},
  {Qian}, {Quirrenbach}, {Wegner}, {Aller}, {Aller}, {Anton}, {Appenzeller},
  {Eckart}, {Kraus}, {Naundorf}, {Kneer}, {Steffen}, \& {Zensus}}]{Wagner96}
{Wagner}, S.~J., {Witzel}, A., {Heidt}, J., {et~al.} 1996, \aj, 111, 2187

\bibitem[{{Woo} \& {Urry}(2002)}]{Woo02}
{Woo}, J.-H., \& {Urry}, C.~M. 2002, \apj, 579, 530

\bibitem[{{Wu} {et~al.}(2012){Wu}, {B{\"o}ttcher}, {Zhou}, {He}, {Ma}, \&
  {Jiang}}]{Wu12}
{Wu}, J., {B{\"o}ttcher}, M., {Zhou}, X., {et~al.} 2012, \aj, 143, 108

\bibitem[{{Wu} {et~al.}(2005){Wu}, {Peng}, {Zhou}, {Ma}, {Jiang}, \&
  {Chen}}]{Wu05}
{Wu}, J., {Peng}, B., {Zhou}, X., {et~al.} 2005, \aj, 129, 1818

\bibitem[{{Wu} {et~al.}(2007){Wu}, {Zhou}, {Ma}, {Wu}, {Jiang}, \&
  {Chen}}]{Wu07}
{Wu}, J., {Zhou}, X., {Ma}, J., {et~al.} 2007, \aj, 133, 1599

\bibitem[{{Xiong} {et~al.}(2020){Xiong}, {Bai}, {Fan}, {Yan}, {Gu}, {Fan},
  {Mao}, {Ding}, {Xue}, \& {Yi}}]{Xiong2020}
{Xiong}, D., {Bai}, J., {Fan}, J., {et~al.} 2020, \apjs, 247, 49

\bibitem[{{Xu} {et~al.}(2019){Xu}, {Hu}, {Webb}, {Bhatta}, {Jiang}, {Chen},
  {Alexeeva}, \& {Li}}]{Xu2019}
{Xu}, J., {Hu}, S., {Webb}, J.~R., {et~al.} 2019, \apj, 884, 92

\bibitem[{{Zhang} {et~al.}(2012){Zhang}, {Dai}, {Wang}, {Zhao}, {Zhang}, \&
  {Cao}}]{Zhang2012}
{Zhang}, B.~K., {Dai}, B.~Z., {Wang}, L.~P., {et~al.} 2012, \mnras, 421, 3111

\bibitem[{{Zhang} {et~al.}(2018){Zhang}, {Wu}, \& {Meng}}]{Zhang18}
{Zhang}, X., {Wu}, J., \& {Meng}, N. 2018, \mnras, 478, 3513

\bibitem[{{Zhang} {et~al.}(2008){Zhang}, {Zheng}, {Zhang}, {Liu}, {Wen}, \&
  {Wang}}]{Zhang08}
{Zhang}, X., {Zheng}, Y.~G., {Zhang}, H.~J., {et~al.} 2008, \aj, 136, 1846

\end{thebibliography}
\bibliographystyle{aasjournal}

\end{document}